\documentclass[12pt]{elsart}
\usepackage{amsmath,amsfonts,amssymb,amsbsy}
\usepackage{natbib}
\usepackage[cp1251]{inputenc}
\usepackage[english]{babel}
\usepackage{graphicx}
\usepackage{color}

\begin{document}
%\setcounter{page}{1}

%\small
\begin{frontmatter}
\title{ 
Accounting for the thickness effect in dynamic spherical indentation of a viscoelastic layer:
Application to non-destructive testing of articular cartilage
}
\author[UK]{I.~Argatov},
%\ead{iva1@aber.ac.uk}
\author[Swiss]{A.U.~Daniels},
\corauth[cor]{Corresponding author.}
\author[UK]{G.~Mishuris\corauthref{cor}},
%\ead{ggm@aber.ac.uk}
\author[Swiss]{S.~Ronken},
\author[Swiss]{D.~Wirz}
\address[UK]{Institute of Mathematics and Physics, Aberystwyth University,
% Ceredigion SY23 3BZ, 
Wales, UK}
\address[Swiss]{Laboratory of Biomechanics \& Biocalorimetry, University Basel, Switzerland}
\begin{abstract}

In recent years, dynamic indentation tests have been shown to be useful both in identification of mechanical properties of biological tissues (such as articular cartilage) and assessing their viability. We consider frictionless flat-ended and spherical sinusoidally-driven indentation tests utilizing displacement-controlled loading protocol. Articular cartilage tissue is modeled as a viscoelastic material with a time-independent Poisson's ratio. We study the dynamic indentation stiffness with the aim of formulating  criteria for evaluation the quality of articular cartilage in order to be able to discriminate its degenerative state. In particular, evaluating the dynamic indentation stiffness at the turning point of the flat-ended indentation test, we introduce the so-called incomplete storage modulus. Considering the time difference between the time moments when the dynamic stiffness vanishes (contact force reaches its maximum) and the dynamic stiffness becomes infinite (indenter displacement reaches its maximum), we introduce the so-called incomplete loss angle. Analogous quantities can be introduced in the spherical sinusoidally-driven indentation test, however, to account for the thickness effect, a special approach is required. 
We apply an asymptotic modeling approach for analyzing and interpreting the results of the dynamic spherical indentation test in terms of the geometrical parameter of the indenter and viscoelastic characteristics of the material. Some implications to non-destructive indentation diagnostics of cartilage degeneration are discussed. 

\end{abstract}

\end{frontmatter}

\setcounter{equation}{0}

\section*{Introduction} 
\label{1DsSectionI}

Joint cartilage is known to have very limited repair capabilities and poorly regenerates. Intensive recent research and development have brought many innovations and also first clinical results in cartilage repair. However, recent clinical, radiological and histological evaluation techniques show somehow contradictory results \citep{Kusano_et_al_2011} and this is why measuring stiffness parameters of cartilage, especially {\it in vivo\/} measurements, are of novel interest nowadays. Cartilage stiffness parameters can be measured in confined \citep{Suh_et_al_1995} or in unconfined \citep{Armstrong_et_al_1984} compression of cartilage specimen. However, both the confined and unconfined compression tests need sample preparation, usually cylindrically shaped specimens of cartilage, and therefore prohibit {\it in vivo\/} measurements. Furthermore, the mapping of the surface is limited by the sample size. These limitations are less restrictive than those usually encountered in indentation testing. 

The first mathematical model allowing to measure stiffness parameters of joint cartilage layer in indentation mode with flat-ended as well as with spherical indenters was developed by \citet{Hayes_et_al_1972}. In the case of a flat-ended indenter of radius $a$ pressed against a sample of thickness $h$, the indentation stiffness defined as the ratio of the contact force $P$ to the indenter displacement $w$ is given by 
\begin{equation}
\frac{P}{w}=\frac{2aE}{1-\nu^2}\kappa_{\rm c}.
\label{1Ds(0.1)}
\end{equation}
Note that compared to \cite{Hayes_et_al_1972}, we replace the shear modulus $G$ with $E/(2(1+nu))$, where $E$ is Young's modulus, $\nu$ is Poisson's ratio. The Hayes model (\ref{1Ds(0.1)}) is based on Hooke's law and takes into account the thickness effect through the stalling factor $\kappa_{\rm c}$. Because the widely used Hayes model assumes linear elasticity, it therefore does not take into consideration the fact that cartilage stiffness parameters are strain-rate dependent, and thus the Hayes model does not consider the dynamic nature of cartilage stiffness. 

Recall that for time-dependent materials \citep{Tschoegl1997}, the dynamic stiffness is characterized by the complex dynamic modulus $E^*=E_1+{\rm i}E_2$ consisting of the storage modulus $E_1$ and the loss modulus $E_2$ with $\rm i$ being the imaginary unit (${\rm i}^2=-1$). On the complex plane, $E_1$, which is a real part of $E^*$, and $E_2$, which is an imaginary part of as imaginary $E^*$, represent the legs (catheti) of a right triangle with the hypotenuse of length $\vert E^*\vert=\sqrt{E_1^2+E_2^2}$. The loss angle, $\delta$, results from the ratio of $E_2$ and $E_1$ through the relationship $\tan\delta=E_2/E_1$.
It should be emphasized that the storage and loss moduli $E_1$ and $E_2$ represent the response of a material to a sinusoidal loading scheme and actually depend on the corresponding angular frequency of sinusoidal oscillations $\omega$. In order to underline this fact we will write $E_1(\omega)$ instead of $E_1$ and so on.

For {\it in vivo\/} (or {\it ex vivo\/}) measurements of cartilage stiffness and mapping a cartilage surface, a mechanical model for cartilage has to be prioritized considering dynamic properties of cartilage and measuring in indentation mode \citep{ Ronken_et_al_2011}. And, application of a single indentation test during arthroscopy allows one to evaluate the quality of cartilage and to detect osteoarthritic degenerative changes \citep{Korhonen_et_al_2003}. The readily available systems on the market for arthroscopic measurements such as the 'Artscan 1000' \citep{Lyyra_et_al_1999,Toyras_et_al_2001} allow only stiffness measurements of the cartilage surface without explicit considering $E_1(\omega)$ and $\delta(\omega)$. 

To ascertain dynamic biomechanical properties of articular cartilage, \citet{Appleyard_et_al_2001} investigated a handheld indentation probe with a flat-ended cylindrical indenter operating in a vibration mode at a single-frequency of 20~Hz. Using the theory of \citet{Hayes_et_al_1972}, the absolute value of the effective complex dynamic modulus can be evaluated as follows:
\begin{equation}
\vert E^*(\omega)\vert=\frac{P_0}{w_0}\frac{(1-\nu^2)}{2a\kappa_{\rm c}}.
\label{1Ds(0.2)}
\end{equation}
Here, $P_0$ is the contact force amplitude, $w_0$ is the displacement amplitude. Note that the complex dynamic modulus $E^*(\omega)$ is termed effective here, because in the case of a poroelastic material such as articular cartilage, the biomechanical response is dependent on the frequency $\omega$ and the boundary conditions for the sample as well. It was also observed \citep{Appleyard_et_al_2001} that when articular cartilage is indented at frequencies above 10~Hz there is marginal change in the effective parameters $\vert E^*(\omega)\vert$ and $\delta(\omega)$ with the effective dynamic modulus being of similar magnitude to the `instanteneous' elastic modulus generated during a rapid load step indentation test. Nevertheless, to the best of our knowledge, no study has analyzed thus far the relationship between the parameters of time-dependent materials measured in a vibration indentation test and in a single indentation test. In order to facilitate such a comparison, we consider sinusoidally-driven displacement-controlled indentation tests. Note that as a first approximation, the half-sinusoidal indentation history can be used for modeling impact tests.

Measuring stiffness parameters of cartilage in indentation mode with spherical tipped indenters has advantages as well as drawbacks. At one hand, with spherical indenters the error obtained when hitting the surface not exactly perpendicular is much smaller than with flat-ended indenters, where the surface is touched with one edge of the indenter first. For example, when the surface with a spherical indenter will be hit with $80^\circ$ instead of $90^\circ$, the result for the stiffness will be underestimated by less than 2\%. On the other hand, spherical indenters underestimate the inhomogeneity and changes in stiffness \citep{Schinagl_et_al_1997} as function of the indentation depth. This is why both theories, for flat-ended and for spherical tipped indenters are provided.

As measuring mechanical properties gained a new importance in recent years, because dynamic indentation tests have been shown to be helpful both in identification of mechanical properties of articular cartilage  and assessing its viability \citep{Bae_et_al_2003,BroomFlachsmann2003}. Indentation stiffness is now accepted as a fundamental indicator of the functional mechanical properties of articular cartilage \citep{Scandiucci_et_al_2006}. 
The dynamic stiffness is defined as the ratio of input force, $P(t)$, to output displacement, $w(t)$. Thus, for a time-dependent material like articular cartilage, the indentation stiffness depends on the indentation protocol, and generally it is a function of time. It is also well known that the indentation stiffness depends on the indenter size as well as on the sample dimensions (see, e.g., Eq.~(\ref{1Ds(0.1)})). This follows from a comparison of the stiffness dimension ${\rm M}{\rm T}^{-2}$ with the dimension ${\rm M}{\rm L}^{-1}{\rm T}^{-2}$ of Young's modulus. From a geometrical point of view, articular cartilage is usually considered as a layer of constant thickness, $h$. In view of the relative mechanical properties of cartilage and subchondral bone, it is assumed that the layer is firmly attached to a non-deformable base. Thus, the indentation scaling factor will depend on the aspect ratio $\alpha=a/h$, where $a$ is the radius if the contact area. 

The above simple analysis is applicable for the linear relationship between the contact force $P(t)$ and the indenter displacement $w(t)$, where the contact radius remains unchanged in time. In spherical indentation, the force-displacement relationship requires a more acute analysis. It will be shown that the results of the dynamic spherical indentation of a time-dependent material depend on the level of indentation.

To a first approximation \citep{HayesMockros1971,ParsonsBlack1977,Lau_et_al_2008}, cartilage tissue can be evaluated mechanically as a viscoelastic material with a time-independent Poisson's ratio, $\nu$, such that the overall constitutive behavior of the material is expressed in terms of its complex modulus $E^*(\omega)$. Indentations tests for viscoelastic materials were studied in a number of publications \citep{Oyen2005,ChengYang2009,ArgatovMishuris2011}. 
We consider flat-ended and spherical indentation tests utilizing displacement-controlled loading protocol with the indenter displacement  modulated according to a sinusoidal law at an angular frequency $\omega=2\pi f$ (rad/s), where $f$ (Hz) is the loading frequency. 
We apply an asymptotic modeling approach for analyzing and interpreting the results of the dynamic spherical indentation test in terms of the geometrical parameter of the indenter (indenter radius, $R$) and viscoelastic characteristics of the material. In particular, we examine the relationships between the storage modulus $E_1(\omega)$ and loss angle $\delta(\omega)$ and the so-called modified storage modulus $\tilde{E}_{3/2}^0(\omega,\varpi_0)$ and the modified loss angle $\tilde{\delta}_{3/2}^0(\omega)$ in the displacement-controlled sinusoidally-driven indentation test.

The rest of the paper is organized as follows. In Section~\ref{1DsSection1}, we consider cylindrical frictionless indentation of a viscoelastic layer. In particular, the linear force-displacement relationship is outlined in Section~\ref{1DsSection1.1}, while the indentation scaling factor for the cylindrical indenter is considered in Section~\ref{1DsSection1.2}. Based on the analogy with the case of harmonic vibrations (considered in Sections~\ref{1DsSection1.3} and \ref{1DsSection1.4}), in \ref{1DsSection1.5}, we introduce the incomplete storage modulus and loss angle as material characteristics that can be assessed directly from a single sinusoidally-driven indentation test. 

In Section~\ref{1DsSection2}, we study spherical indentation of a viscoelastic layer. Based on the general solution obtained by \citet{Ting1968}, in Sections~\ref{1DsSection2.1} and \ref{1DsSection2.2}, we write out the force-displacement relationship for the loading and unloading stages, respectively. The indentation scaling factor for the spherical indenter is introduced in Section~\ref{1DsSection2.3}. In Section~\ref{1DsSection2.4}, we introduce the so-called modified incomplete storage modulus and loss angle, and investigate their behavior in Section~\ref{1DsSection2.5} for the standard viscoelastic solid model.

In Section~\ref{1DsSection3}, we actually consider the thickness effect in spherical indentation of a viscoelastic layer. By analogy with the elastic case, we introduce the quantity $\tilde{E}_{3/2}^0(\omega,\varpi_0)$ (which is called the modified storage modulus, in view of its relation to the storage modulus $E_1(\omega)$) while the modified loss angle $\tilde{\delta}_{3/2}^0(\omega)$ is introduced according to a standard interpretation of the time lag between the peak force and peak displacement. Some properties of these parameters, which turned out to be dependent on the level of indentation, are illustrated for the standard viscoelastic solid model. Low- and high-frequency asymptotic analysis of the quantities $\tilde{E}_\beta^0(\omega,\varpi_0)$ and $\tilde{\delta}_\beta^0(\omega)$ is presented in Sections~\ref{1DsSection3.2} and \ref{1DsSection3.3}, respectively.

Finally, in Sections~\ref{1DsSectionD} and \ref{1DsSectionC}, we outline a discussion of the results obtained and formulate our conclusions.

\section{Cylindrical frictionless indentation of a viscoelastic layer}
\label{1DsSection1}

\subsection{Liner force-displacement relationship}
\label{1DsSection1.1}

We consider a viscoelastic layer bonded to a rigid substrate indented by a flat-ended cylindrical indenter. 
For the sake of simplicity, we neglect friction and assume that Poisson's ratio, $\nu$, of the layer material is time independent. Then, applying the elastic-viscoelastic correspondence principle \citep{Christensen1971}, one can arrive at the following equation between the applied force $P(t)$ and the displacement of the indenter $w(t)$ \citep{ZhangZhang2004,Cao_et_al_2009}:
\begin{equation}
P(t)=\frac{2a}{1-\nu^2}\,\kappa_{\rm c}(\alpha)
\int\limits_{0-}^t E(t-\tau)\frac{dw}{d\tau}(\tau)\,d\tau.
\label{1Ds(1.1)}
\end{equation}
Here, $a$ is the radius of the contact area, $h$ is the layer thickness, $E(t)$ is the relaxation modulus, $t$ is the time variable, $t=0-$ is the time moment just preceding the initial moment of contact, $\kappa_{\rm c}(\alpha)$ is a dimensionless factor, which is determined from the solution of elastic contact problem for a cylindrical indenter, $\alpha$ is the relative radius of the contact area that is, i.\,e.,
\begin{equation}
\alpha=\frac{a}{h}.
\label{1Ds(1.2a)}
\end{equation}

\begin{figure}[h!]
%\vskip-1.0cm    
    \centering
    \hbox{
    \includegraphics[scale=0.32]{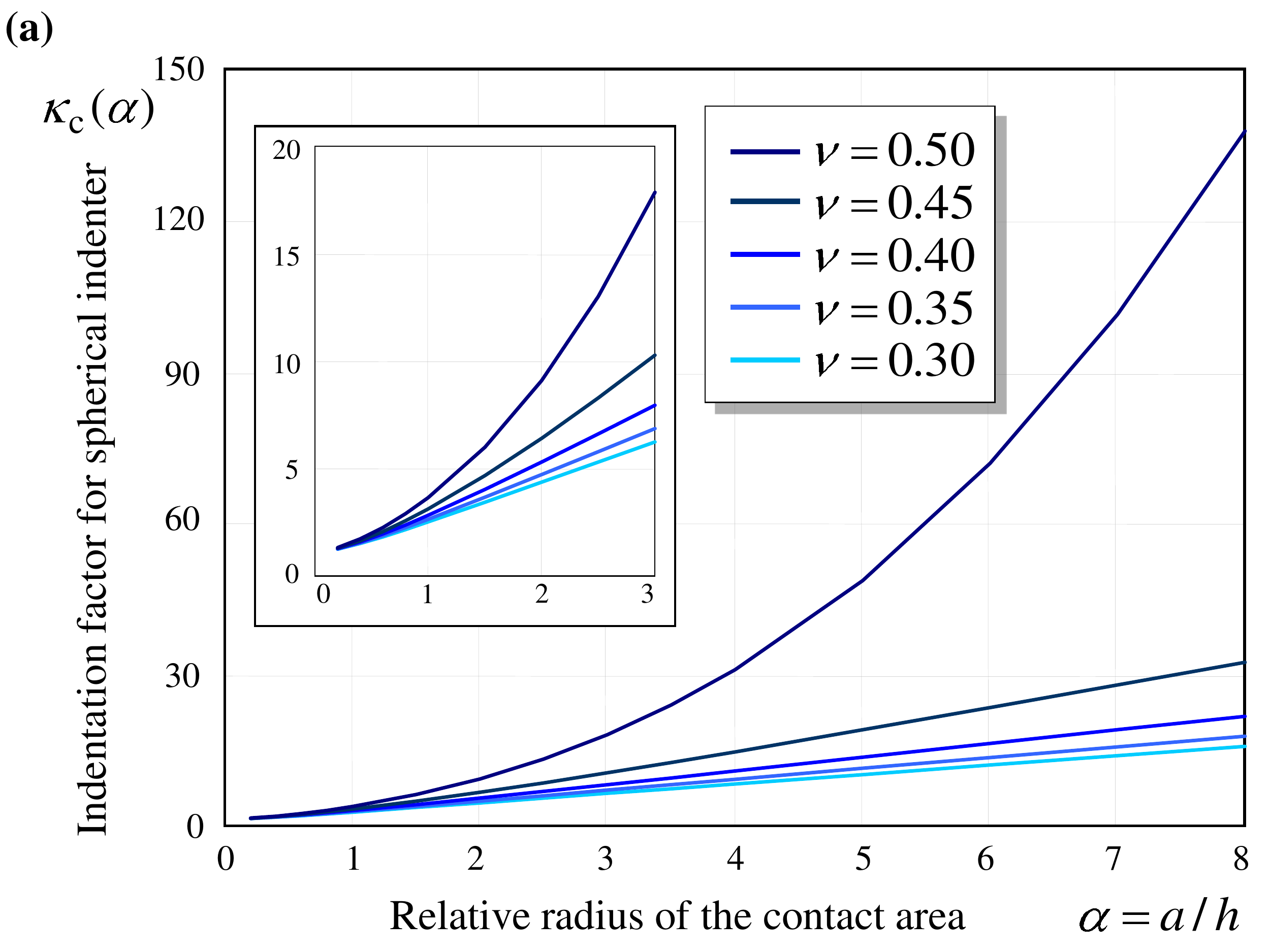}%\hskip-0.3cm
    \includegraphics[scale=0.32]{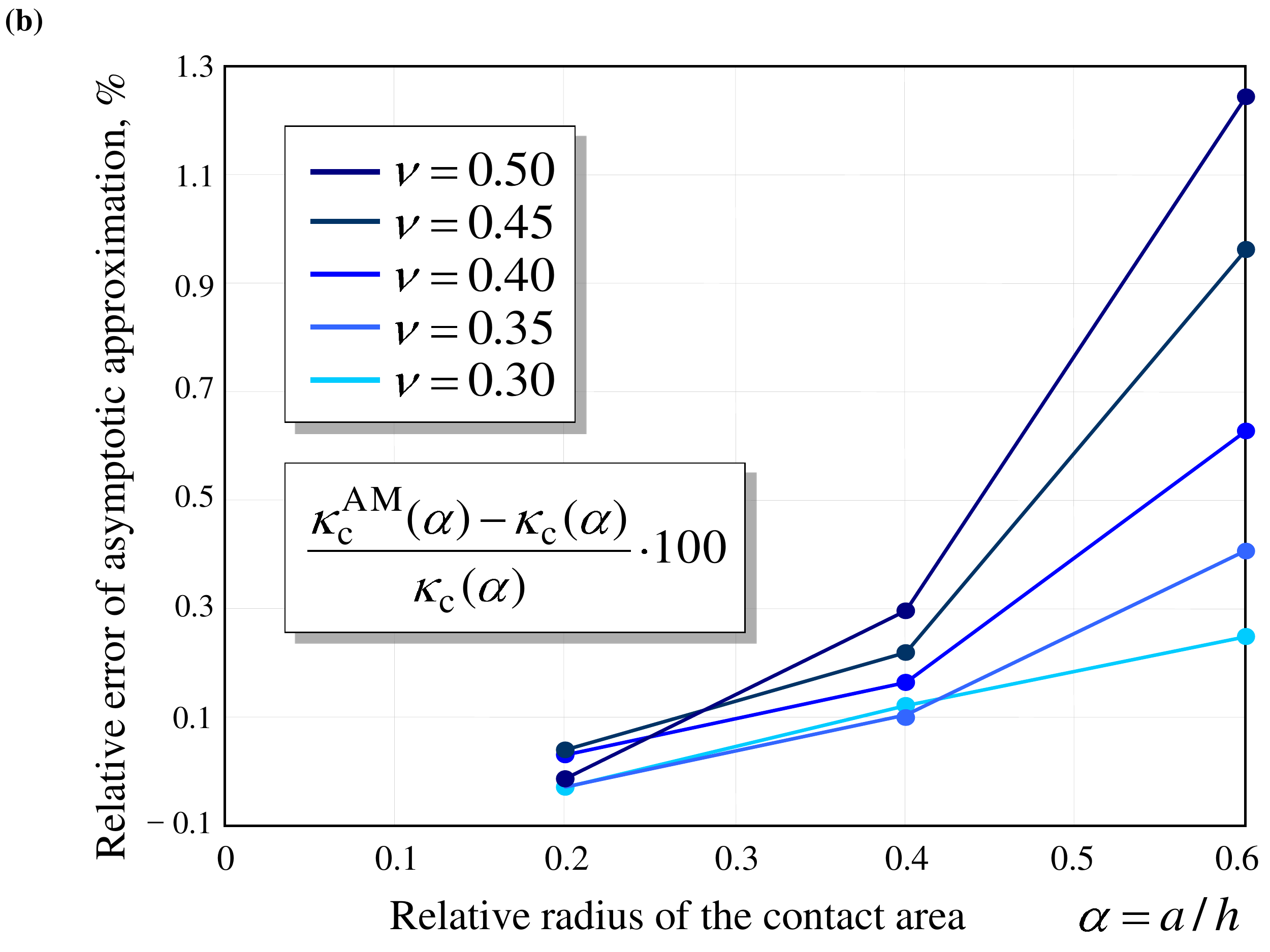}
    }
%\vskip-3.5cm    
    \caption{(a) Indentation scaling factor for the cylindrical indenter as a function of relative contact radius; 
    (b) Relative error of of the asymptotic approximation (\ref{1Ds(1a.1)}).    }
%\vskip-1.0cm        
    \label{Fig-kappa-c}
\end{figure}

Note that the dependence of $\kappa_{\rm c}(\alpha)$ on Poisson's ratio is not indicated explicitly. Fig.\,\ref{Fig-kappa-c}a illustrates the behavior of $\kappa_{\rm c}(\alpha)$ for different values of $\nu$ based on the numerical solution obtained by \citet{Hayes_et_al_1972}.

Denoting $E(t)=E_\infty\Psi(t)$, where $E_\infty$ is the relaxed elastic modulus (the limit of modulus $E(t)$ at $t\to\infty$), $\Psi(t)$ is the relaxation function, we rewrite Eq.~(\ref{1Ds(1.1)}) in the following form:
\begin{equation}
P(t)=\frac{2aE_\infty}{1-\nu^2}\,\kappa_{\rm c}(\alpha)
\int\limits_{0-}^t \Psi(t-\tau)\frac{dw}{d\tau}(\tau)\,d\tau.
\label{1Ds(1.2)}
\end{equation}

Inverting the relationship (\ref{1Ds(1.2)}), we obtain
\begin{equation}
w(t)=\frac{1-\nu^2}{2aE_\infty}\,\frac{1}{\kappa_{\rm c}(\alpha)}
\int\limits_{0-}^t \Phi(t-\tau)\frac{dP}{d\tau}(\tau)\,d\tau,
\label{1Ds(1.3)}
\end{equation}
where $\Phi(t)$ is the creep function.

\subsection{Indentation scaling factor for the cylindrical indenter}
\label{1DsSection1.2}

According to \citet{Vorovich_et_al_1974,Argatov2002}, the following asymptotic model takes place for the indentation scaling factor $\kappa_{\rm c}(\alpha)$:
\begin{eqnarray}
\kappa_{\rm c}(\alpha) & = & 1+\alpha\frac{2a_0}{\pi}+
\alpha^2\Bigl(\frac{2a_0}{\pi}\Bigr)^2+
\alpha^3\biggl[\Bigl(\frac{2a_0}{\pi}\Bigr)^3+\frac{8 a_1}{3\pi}\biggr]
\nonumber \\
{} & {} & {}+\alpha^4\biggl[\Bigl(\frac{2a_0}{\pi}\Bigr)^4+\frac{32 a_0 a_1}{3\pi^2}\biggr]
+O(\alpha^5).
\label{1Ds(1a.1)}
\end{eqnarray}
Here, $a_0$ and $a_1$ are asymptotic constants depending on Poisson's ratio $\nu$ given by
$$
a_m=\frac{(-1)^m}{2^{2m}(m!)^2}\int\limits_0^\infty
\bigl[1-\mathcal{L}(\lambda)\bigr]\lambda^{2m}\,d\lambda.
$$
In the case of a layer bonded to a rigid base, we have
$$
\mathcal{L}(\lambda)=\frac{2\varkappa\, {\rm sh\,}2\lambda-4\lambda}{2\varkappa\,{\rm ch\,}2\lambda+1+\varkappa^2+4\lambda^2},
$$
where $\varkappa=3-4\nu$ is Kolosov's constant. 

To determine the range of validity of the asymptotic model (\ref{1Ds(1a.1)}), we compare its predictions with the numerical solution given by \citet{Hayes_et_al_1972}. As it could be expected (see Fig.\,\ref{Fig-kappa-c}b), the accuracy of the approximation $\kappa_{\rm c}^{\rm AM}(\alpha)$ given by (\ref{1Ds(1a.1)}) decreases as Poisson's ratio approaches $0{.}5$ .
Fig.\,\ref{Fig-kappa-c}b shows that asymptotic approximation (\ref{1Ds(1a.1)}) is quite accurate in the range $\alpha\in(0,0{.}6)$, that is for the indenter diameter less than the layer thickness. Note here that the substrate effect on the incremental indentation stiffness was considered in the elastic case in \citep{Argatov2010}.

\subsection{Harmonic vibration}
\label{1DsSection1.3}

Observe that Eq.~(\ref{1Ds(1.1)}) assumes that the layer material was at rest for $t<0$. In order to study harmonic vibrations of a viscoelastic layer, we should replace Eq.~(\ref{1Ds(1.1)}) with the following one:
\begin{equation}
P(t)=\frac{2a}{1-\nu^2}\,\kappa_{\rm c}(\alpha)
\int\limits_{-\infty}^t E(t-\tau)\frac{dw}{d\tau}(\tau)\,d\tau.
\label{1Ds(1.4)}
\end{equation}

Substituting a harmonic displacement 
$w(t)={\rm Im}\{w_0\exp({\rm i}\omega t)\}$ with amplitude $w_0$ and frequency $\omega$ into Eq.~(\ref{1Ds(1.4)}), one can arrive at the following equation:
\begin{equation}
P(t)=\frac{2aw_0}{1-\nu^2}\,\kappa_{\rm c}(\alpha)
{\,\rm Im}\{E^*(\omega)\exp({\rm i}\omega t)\}.
\label{1Ds(1.5)}
\end{equation}
Here, $\rm Im$ denotes the imaginary part of a complex number, $E^*(\omega)$ is the complex relaxation modulus given by
\begin{equation}
E^*(\omega)={\rm i}\omega
\int\limits_0^{\infty} E(s)\exp(-{\rm i}\omega s)\,ds.
\label{1Ds(1.6)}
\end{equation}

By convention \citep{Pipkin1986,Tschoegl1997}, we define the storage modulus, $E_1(\omega)$, and the loss modulus, $E_2(\omega)$, as the real and imaginary parts of $E^*(\omega)$, respectively, i.\,e.,
\begin{equation}
E^*(\omega)=E_1(\omega)+{\rm i}E_2(\omega).
\label{1Ds(1.7)}
\end{equation}

From Eqs.~(\ref{1Ds(1.6)}) and (\ref{1Ds(1.7)}), it follows that
\begin{equation}
E_1(\omega)=\omega E_\infty
\int\limits_0^{\infty} \Psi(s)\sin\omega s\,ds,
\label{1Ds(1.8)}
\end{equation}
\begin{equation}
E_2(\omega)=\omega E_\infty
\int\limits_0^{\infty} \Psi(s)\cos\omega s\,ds.
\label{1Ds(1.9)}
\end{equation}

Furthermore, according to Eq.~(\ref{1Ds(1.5)}), we can write
\begin{equation}
P(t)=P_0\sin(\omega t+\delta),
\label{1Ds(1.10)}
\end{equation}
where $P_0$ is the force amplitude, $\delta$ is the phase angle between the harmonic displacement and the force, given by the formulas 
\begin{equation}
P_0=\frac{2a}{1-\nu^2}\,\kappa_{\rm c}(\alpha)
\vert E^*(\omega)\vert w_0,
\label{1Ds(1.11)}
\end{equation}
\begin{equation}
\cos\delta=\frac{E_1(\omega)}{\vert E^*(\omega)\vert},\quad
\sin\delta=\frac{E_2(\omega)}{\vert E^*(\omega)\vert},\quad
\vert E^*(\omega)\vert=\sqrt{E_1(\omega)^2+E_2(\omega)^2}.
\label{1Ds(1.12)}
\end{equation}

Observe that the phase angle $\delta$ depends on the frequency $\omega$ (this is not indicated in notation for simplicity). 

Finally, note that the vibration indentation tests should be accomplished with a quasistatic preload to ensure a complete contact between the indenter's base and the layer surface, since tensile stresses are not allowed in frictionless indentation.  

\subsection{Determination of the complex relaxation modulus via vibration indentation tests}
\label{1DsSection1.4}

We assume that the displacement and force amplitudes $w_0$ and $P_0$ as well as the phase angle $\delta$ are experimentally measurable quantities. Then, Eqs.~(\ref{1Ds(1.11)}) and (\ref{1Ds(1.12)}) yield the following equations \citep{Cao_et_al_2009}:
\begin{equation}
E_1(\omega)=\frac{1-\nu^2}{2a\kappa_{\rm c}(\alpha)}\frac{P_0}{w_0}\cos\delta,
\label{1Ds(1.13)}
\end{equation}
\begin{equation}
E_2(\omega)=\frac{1-\nu^2}{2a\kappa_{\rm c}(\alpha)}\frac{P_0}{w_0}\sin\delta.
\label{1Ds(1.14)}
\end{equation}

Thus, for a given constant frequency $\omega$, the vibration indentation test yields the storage and loss moduli $E_1(\omega)$ and $E_2(\omega)$, if the amplitude ratio $P_0/w_0$ and the phase angle $\delta$ are known from the experiment. 

Further, let $t_m$ denote the moment of time when the indentation speed $\dot{w}(t)$ vanishes, that is, when
$\dot{w}(t_m)=0$ and $t_m=\pi/(2\omega)+\pi k/\omega$, $k=0,\pm 1,\pm 2,\ldots\,$.
Considering the indentation process over a half of period $t\in(0,\pi/\omega)$, we will have
$t_m=\pi/(2\omega)$ and, correspondingly, $w(t_m)=w_0$ and $P(t_m)=P_0\cos\delta$. Hence, taking Eq.~(\ref{1Ds(1.13)}) into account, we obtain the formula 
\begin{equation}
E_1(\omega)=\frac{1-\nu^2}{2a\kappa_{\rm c}(\alpha)}\frac{P(t_m)}{w(t_m)},
\label{1Ds(1.15)}
\end{equation}
where $t_m$ is a time moment such that $\dot{w}(t_m)=0$.

\begin{figure}[h!]
%\vskip-1.0cm    
    \centering
    \hbox{
    \includegraphics[scale=0.29]{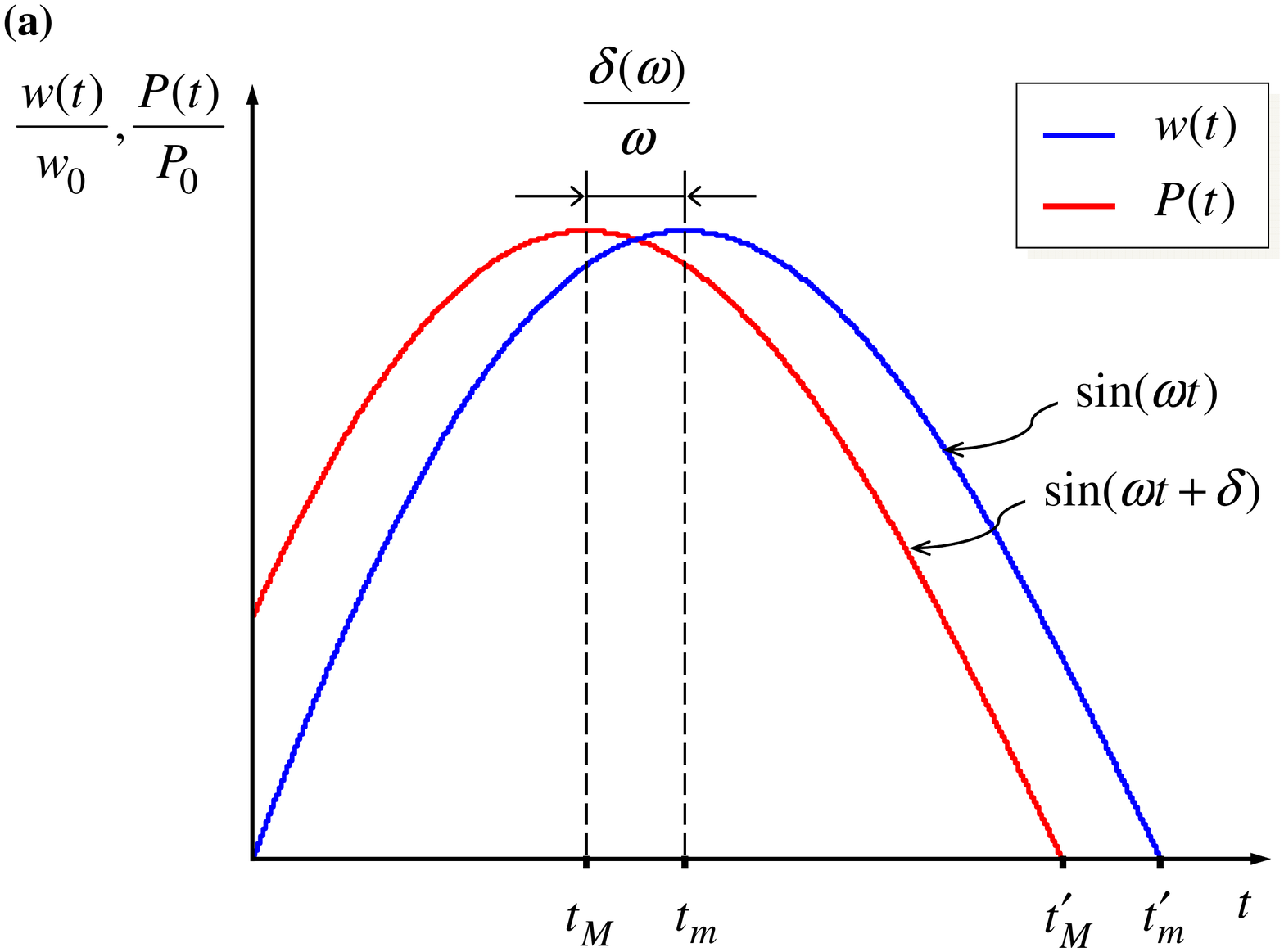}%\hskip-0.3cm
    \includegraphics[scale=0.29]{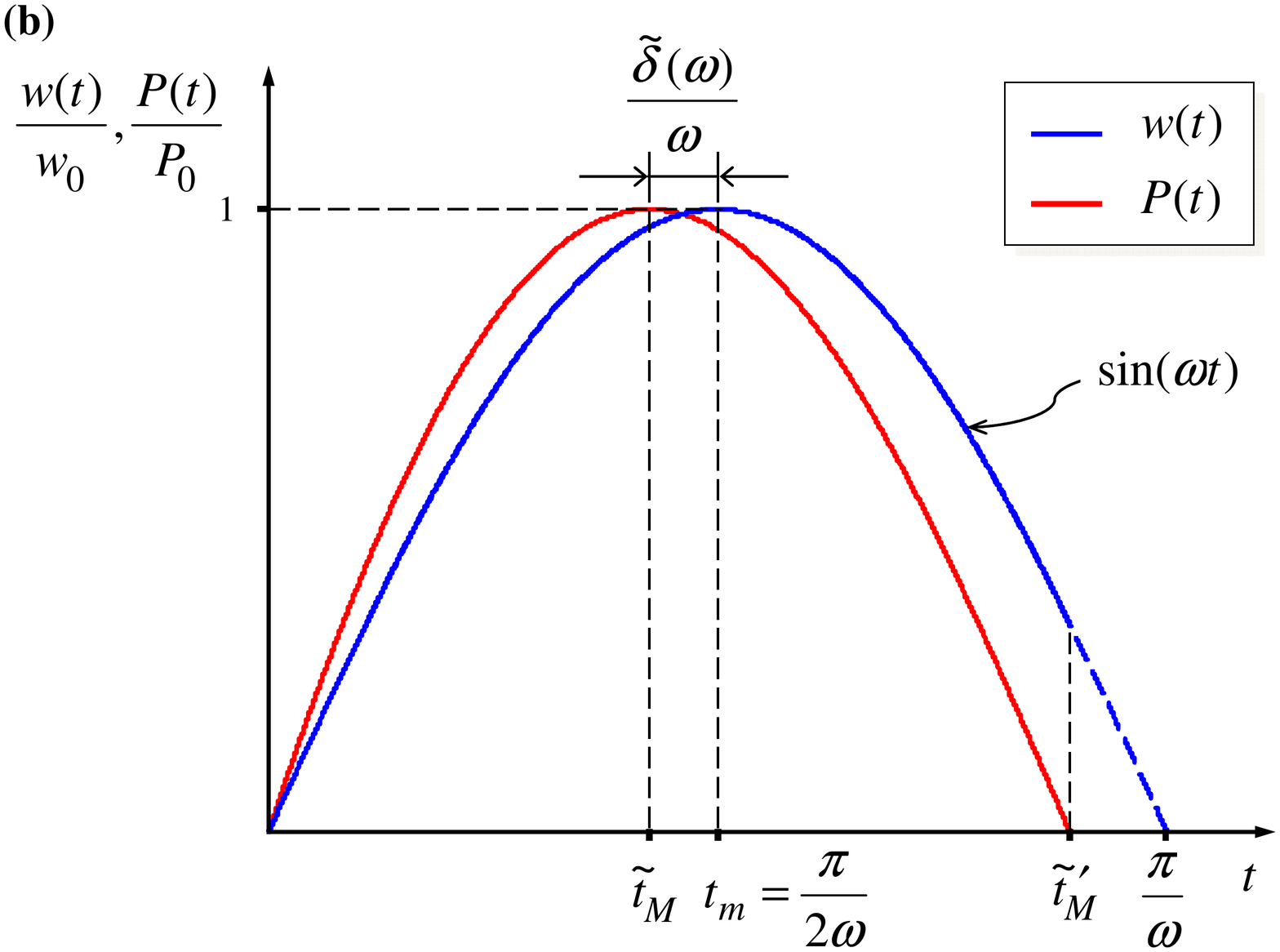}
    }
%\vskip-3.5cm    
    \caption{(a) Displacement-controlled oscillation test; 
    (b) Displacement-controlled indentation test.    }
%\vskip-1.0cm        
    \label{Fig-test}
\end{figure}

Thus, according to Eq.~(\ref{1Ds(1.15)}), the ratio $P(t_m)/w(t_m)$ at the time moment of the displacement extremum determines the storage modulus. 

Let now $t_m^\prime$ be the moment of time when the indentation displacement $w(t)$ vanishes, i.\,e.,
$w(t_m^\prime)=0$ and $t_m^\prime=\pi k/\omega$, $k=0,\pm 1,\pm 2,\ldots\,$. (Note that for harmonic vibrations
$t_m=t_m+\pi/(2\omega)$.) Taking into account Eq.~(\ref{1Ds(1.14)}), we obtain
\begin{equation}
E_2(\omega)=\frac{1-\nu^2}{2a\kappa_{\rm c}(\alpha)}\frac{\omega P(t_m^\prime)}{\dot{w}(t_m^\prime)},
\label{1Ds(1.16)}
\end{equation}
where $\dot{w}(t_m^\prime)$ is the indentation speed when the indentation displacement vanishes.

{\rem{
For the sake of completeness, we provide below the dual-conjugate formulas for Eqs.~{\rm(\ref{1Ds(1.15)})} and {\rm(\ref{1Ds(1.16)})}. Let $t_M$ denote the moment of time when the derivative of the contact force, $\dot{P}(t)$, vanishes, that is, when
$\dot{P}(t_M)=0$ and $t_M=(\pi/2-\delta)/\omega+\pi k/\omega$, $k=0,\pm 1,\pm 2,\ldots\,$.
Let also $t_M^\prime$ be the time moment when the contact force $P(t)$ vanishes, i.\,e.,
$P(t_M^\prime)=0$ and $t_M^\prime=(\pi-\delta)/\omega+\pi k/\omega$, $k=0,\pm 1,\pm 2,\ldots\,$. 
Then, according to Eqs.~{\rm(\ref{1Ds(1.13)})} and {\rm(\ref{1Ds(1.14)})}, the following relationships hold true:
\begin{equation}
\frac{E_1(\omega)}{\vert E^*(\omega)\vert^2}
=\frac{2a\kappa_{\rm c}(\alpha)}{1-\nu^2}
\frac{w(t_M)}{P(t_M)},
\label{1Ds(1.17)}
\end{equation}
\begin{equation}
\frac{E_2(\omega)}{\vert E^*(\omega)\vert^2}
=-\frac{2a\kappa_{\rm c}(\alpha)}{1-\nu^2}
\frac{\omega w(t_M^\prime)}{\dot{P}(t_M^\prime)}.
\label{1Ds(1.18)}
\end{equation}
Recall that the magnitude of the complex modulus, $\vert E^*(\omega)\vert$, is determined by the last formula {\rm(\ref{1Ds(1.12)})}.
}}

\subsection{Indentation test with a sinusoidal displacement. Incomplete storage modulus and loss angle}
\label{1DsSection1.5}

Let us first consider a single indentation test with a prescribed sinusoidal displacement according to the law
\begin{equation}
w(t)=w_0\sin\omega t, \quad t\in(0,\pi/\omega).
\label{1Ds(4.1)}
\end{equation}
Here, $w_0$ is the maximum depth of indentation, $\omega$ is a given quantity having the dimension of reciprocal time. The quantity 
\begin{equation}
t_m=\frac{\pi}{2\omega}
\label{1Ds(4.1a)}
\end{equation} has a physical meaning of the time moment when the indentation displacement reaches its maximum. We emphasize that due to viscoelastic properties of the layer material, the duration of contact will be less than $\pi/\omega$.

According to Eq.~(\ref{1Ds(1.1)}), we get
\begin{equation}
P(t_m)=\frac{2a}{1-\nu^2}\,\kappa_{\rm c}(\alpha)\omega w_0
\int\limits_{0}^{t_m} E(t_m-\tau)\cos\omega\tau\,d\tau,
\label{1Ds(4.2)}
\end{equation}
where $t_m=\pi/(2\omega)$.

By analogy with Eq.~(\ref{1Ds(1.15)}), we define
\begin{equation}
\tilde{E}_1(\omega)=\frac{1-\nu^2}{2a\kappa_{\rm c}(\alpha)}\frac{P(t_m)}{w(t_m)},
\label{1Ds(4.3)}
\end{equation}
where $w(t_m)=w_0$ (see Eq.~(\ref{1Ds(4.1)})).

It is clear that the quantity $\tilde{E}_1(\omega)$, introduced for single indentation test, differs from the storage modulus $E_1(\omega)$, introduced for vibration indentation test.

In view of (\ref{1Ds(4.2)}), Eq.~(\ref{1Ds(4.3)}) yields
\begin{equation}
\tilde{E}_1(\omega) = \omega 
\int\limits_{0}^{\pi/(2\omega)} E(s)\sin\omega s\,ds.
\label{1Ds(4.4)}
\end{equation}

Recalling the notation $E(t)=E_\infty\Psi(t)$, we rewrite Eq.~(\ref{1Ds(4.4)}) in the form
\begin{equation}
\tilde{E}_1(\omega)= \omega E_\infty
\int\limits_{0}^{\pi/(2\omega)} \Psi(s)\sin\omega s\,ds.
\label{1Ds(4.5)}
\end{equation}
Comparing Eqs.~(\ref{1Ds(1.8)}) and (\ref{1Ds(4.5)}), we see that their right-hand sides differ only by the integral upper limits. 

Now, let $\tilde{t}_M$ be the time moment when the contact force (\ref{1Ds(1.1)}) corresponding to the indentation law (\ref{1Ds(4.1)}) reaches its maximum, i.\,e., $\dot{P}(\tilde{t}_M)=0$. Then, by analogy with the case of linear harmonic vibrations (see Eq.~(\ref{1Ds(1.10)})), we put
\begin{equation}
\tilde{\delta}(\omega)=\frac{\pi}{2}-\omega \tilde{t}_M.
\label{1Ds(4.6)}
\end{equation}
The quantity $\tilde{\delta}(\omega)$ is called the incomplete loss angle determined from the sinusoidally-driven displacement-controlled cylindrical indentation test \citep{Argatov2012}. 

In view of (\ref{1Ds(4.1a)}), formula (\ref{1Ds(4.6)}) can be rewritten as follows:
\begin{equation}
\tilde{\delta}(\omega)=\frac{\pi}{2}\frac{(t_m- \tilde{t}_M)}{t_m}.
\label{1Ds(4.7)}
\end{equation}

The interrelations between the quantities $\tilde{E}_1(\omega)$, $\tilde{\delta}(\omega)$
and $E_1(\omega)$, $\delta(\omega)$ were investigated in \citep{Argatov2012}. It was shown that they asymptotically coincide, respectively, in both the low and high frequency limits, while within the intermediate range of $\omega$, the differences depend on the viscoelastic model in question, that is on the properties of the relaxation modulus $E(t)$. 

\section{Spherical frictionless indentation of a viscoelastic layer}
\label{1DsSection2}

\subsection{Force-displacement relationship in the loading stage}
\label{1DsSection2.1}

Applying the general solution obtained by \citet{Ting1968} for a class of viscoelastic contact problems in terns of the corresponding elastic solutions, we will have
\begin{equation}
P(t)=\frac{4E_\infty h^3}{3(1-\nu^2)R}
\int\limits_{0}^t\frac{d}{d\tau}\bigl\{
\alpha(\tau)^3\mathcal{F}(\alpha(\tau))\bigr\}\Psi(t-\tau)\,d\tau,
\label{1Ds(2.2)}
\end{equation}
\begin{equation}
w(t)=\frac{h^2\alpha(t)^2}{R}\,\mathcal{G}(\alpha(t)).
\label{1Ds(2.3)}
\end{equation}
Here, $\alpha(t)$ is the variable relative radius of the contact area, i.\,e. (cf. Eq.~(\ref{1Ds(1.2a)}))
\begin{equation}
\alpha(t)=\frac{a(t)}{h},
\label{1Ds(2.3a)}
\end{equation}
while $\mathcal{F}(\alpha)$ and $\mathcal{G}(\alpha)$ are depending on Poisson's ratio $\nu$ dimensionless factors such that $\mathcal{F}(0)=\mathcal{G}(0)=1$.

We will use formulas (\ref{1Ds(2.2)}) and (\ref{1Ds(2.3)}) under the assumption that the relative contact radius $\alpha(t)$ monotonically increases in the time interval $(0,t_m)$, where $t_m$ is a certain moment of time.

\subsection{Force-displacement relationship in the unloading stage}
\label{1DsSection2.2}

Let us assume that the relative contact radius $\alpha(t)$ decreases to zero in the interval $t\in(t_m,t_c)$, where $t_c$ is the time of contact of the indenter with the layer surface. Once again, making use of the general solution derived by \citet{Ting1968} for the case when the variation of contact radius posses a single maximum, we obtain
\begin{equation}
P(t)=\frac{4E_\infty h^3}{3(1-\nu^2)R}
\int\limits_{0-}^{t_1(t)}\frac{d }{d\tau}\bigl\{
\alpha(\tau)^3\mathcal{F}(\alpha(\tau))\bigr\}\Psi(t-\tau)\,d\tau,
\label{1Ds(2.4)}
\end{equation}
\begin{eqnarray}
w(t) & = & \frac{h^2}{R}\Biggl\{ \alpha(t)^2\mathcal{G}(\alpha(t))
\nonumber \\
{} & {} & {}-\int\limits_{t_m}^{t}\Phi(t-\tau)\frac{\partial}{\partial\tau}
   \int\limits_{t_1(\tau)}^{\tau}\frac{d }{d\eta}\bigl\{
\alpha(\eta)^2\mathcal{G}(\alpha(\eta))\bigr\}\Psi(\tau-\eta)\,d\eta d\tau
\Biggr\}.
\label{1Ds(2.5)}
\end{eqnarray}
Here, $t_1(\tau)$ is the time moment prior to $t_m$ such that the contact radius $a(\tau)$ is equal to the prior contact radius $a(t_1(\tau))$. The function $t_1(t)$ remains to be calculated. 

If the indenter displacement $w(t)$ is prescribed, taking into account the relation 
$a(t_1)=a(t)$ for $t_1\leq t_m\leq t$, we arrive at the equation 
\begin{equation}
w(t) = \frac{h^2\alpha(t)^2}{R}\mathcal{G}(\alpha(t)),
\label{1Ds(2.6)}
\end{equation}
from which we get
\begin{equation}
t_1(t) = w^{-1}\Bigl(
\frac{h^2\alpha(t)^2}{R}\mathcal{G}(\alpha(t))\Bigr),\quad t\in[t_m,t_c].
\label{1Ds(2.7)}
\end{equation}

Finally, note that in the inner integral in (\ref{1Ds(2.5)}), for $\eta$ such that 
$t_1(\tau)\leq \eta\leq t_m$, the function $\alpha(\eta)$  should be calculated according to Eqs.~(\ref{1Ds(2.2)}) and (\ref{1Ds(2.3)}).

\subsection{Indentation scaling factor for the spherical indenter}
\label{1DsSection2.3}

In the elastic case, according to the notation used in Eqs.~(\ref{1Ds(2.2)}) and (\ref{1Ds(2.3)}), we have
\begin{equation}
P=\frac{4E h^3}{3(1-\nu^2)R}\alpha^3\mathcal{F}(\alpha),
\label{1Ds(2.10)}
\end{equation}
\begin{equation}
w=\frac{h^2}{R}\,\alpha^2\mathcal{G}(\alpha).
\label{1Ds(2.11)}
\end{equation}
Here, $\alpha$ is the relative radius of the contact area as defined by formula (\ref{1Ds(1.2a)}).

Representing Eq.~(\ref{1Ds(2.11)}) in the form
\begin{equation}
\frac{\sqrt{wR}}{h}=\alpha\sqrt{\mathcal{G}(\alpha)},
\label{1Ds(2.12)}
\end{equation}
and taking into account that $\mathcal{G}(0)=1$, we see that Eq.~(\ref{1Ds(2.11)}) can be inverted as
\begin{equation}
\alpha=\frac{\sqrt{wR}}{h}g(\varpi),
\label{1Ds(2.13)}
\end{equation}
where
\begin{equation}
\varpi=\frac{\sqrt{wR}}{h}.
\label{1Ds(2.14)}
\end{equation}

Substituting the expression (\ref{1Ds(2.13)}) into Eq.~(\ref{1Ds(2.10)}), we obtain the following relationship:
\begin{equation}
P=\frac{4E\sqrt{R}}{3(1-\nu^2)}w^{3/2}f(\varpi).
\label{1Ds(2.15)}
\end{equation}
Here we introduced the notation
\begin{equation}
f(\varpi)=g(\varpi)^3 \mathcal{F}(\varpi g(\varpi)).
\label{1Ds(2.16)}
\end{equation}
It is clear that $f(0)=g(0)=1$.

Finally, in view of Eq.~(\ref{1Ds(2.12)}), we can represent Eq.~(\ref{1Ds(2.15)}) as
\begin{equation}
P=\frac{4E\sqrt{R}}{3(1-\nu^2)}w^{3/2}\kappa_{\rm s}(\alpha),
\label{1Ds(2.17)}
\end{equation}
where we introduced the indentation scaling factor
\begin{equation}
\kappa_{\rm s}(\alpha)=f\bigl(\alpha\sqrt{\mathcal{G}(\alpha)}\bigr).
\label{1Ds(2.18)}
\end{equation}
We emphasize that $\kappa_{\rm s}(\alpha)$ is normalized in such a way that $\kappa_{\rm s}(0)=1$. Numerical values for $\kappa_{\rm s}(\alpha)$ for a range of parameters $\alpha$ and $\nu$ are given in 
Table~\ref{table:kappa-s} based on the results obtained by \citet{Hayes_et_al_1972}. We note that 
$\kappa_{\rm s}(\alpha)=(3/2)\sqrt{\chi}\kappa$, where $\chi$ and $\kappa$ are parameters employed in their analysis. 

\begin{table}[!h]
\caption{Values of $\kappa_{\rm s}(\alpha)$ for the spherical indenter.}    
\vskip0.2cm
\begin{center}
\begin{tabular}{c|c|c|c|c|c}\hline
$a/h$ & $\nu=0{.}30$ & $\nu=0{.}35$ & $\nu=0{.}40$ & $\nu=0{.}45$ & $\nu=0{.}50$  \\ \hline
0.04 &	1.034 &	1.035 &	1.037 &	1.040 &	1.044 \\ 
0.06 &	1.052 &	1.055 &	1.058 &	1.063 &	1.069 \\ 
0.08 &	1.072 &	1.075 &	1.080 &	1.086 &	1.094 \\ 
0.1 &	1.091 &	1.095 &	1.102 &	1.109 &	1.120 \\ 
0.2 &	1.197 &	1.208 &	1.221 &	1.240 &	1.266 \\ 
0.3 &	1.315 &	1.333 &	1.356 &	1.389 &	1.435 \\ 
0.4 &	1.445 &	1.472 &	1.507 &	1.557 &	1.629 \\ 
0.5 &	1.585 &	1.622 &	1.672 &	1.744 &	1.849 \\ 
0.6 &	1.734 &	1.784 &	1.851 &	1.949 &	2.096 \\ 
0.7 &	1.892 &	1.955 &	2.044 &	2.171 &	2.370 \\ 
0.8 &	2.058 &	2.135 &	2.245 &	2.410 &	2.671 \\ 
0.9 &	2.228 &	2.322 &	2.459 &	2.664 &	2.998 \\ 
1.0 &	2.405 &	2.517 &	2.681 &	2.932 &	3.354 \\ 
1.25 &	2.863 &	3.023 &	3.268 &	3.659 &	4.359 \\ 
1.5 &	3.337 &	3.554 &	3.891 &	4.455 &	5.532 \\ 
1.75 &	3.821 &	4.100 &	4.542 &	5.311 &	6.886 \\ 
2.0 &	4.311 &	4.654 &	5.212 &	6.221 &	8.427 \\ 
2.25 &	4.808 &	5.217 &	5.903 &	7.179 &	10.163 \\ 
2.5 &	5.309 &	5.787 &	6.604 &	8.180 &	12.113 \\ 
2.75 &	5.810 &	6.363 &	7.317 &	9.215 &	14.288 \\ 
3.0 &	6.316 &	6.939 &	8.040 &	10.289 &	16.699 \\ 
 \hline
\end{tabular}    
\end{center}
\label{table:kappa-s}
\end{table}

According to \citet{Argatov2001}, the following asymptotic expansion holds true:
\begin{eqnarray}
f(\varpi) & = & 1+\varpi\frac{2a_0}{\pi}+
\varpi^2\frac{14a_0^2}{3\pi^2}+
\varpi^3\biggl[\frac{320a_0^3}{27\pi^3}+\frac{32 a_1}{15\pi}\biggr]
\nonumber \\
{} & {} & {}+\varpi^4\biggl[\frac{286a_0^4}{9\pi^4}+\frac{64 a_0 a_1}{5\pi^2}\biggr]
+O(\varpi^5).
\label{1Ds(2.19)}
\end{eqnarray}

\begin{figure}[h!]
%\vskip-1.0cm    
    \centering
    \hbox{
    \includegraphics[scale=0.32]{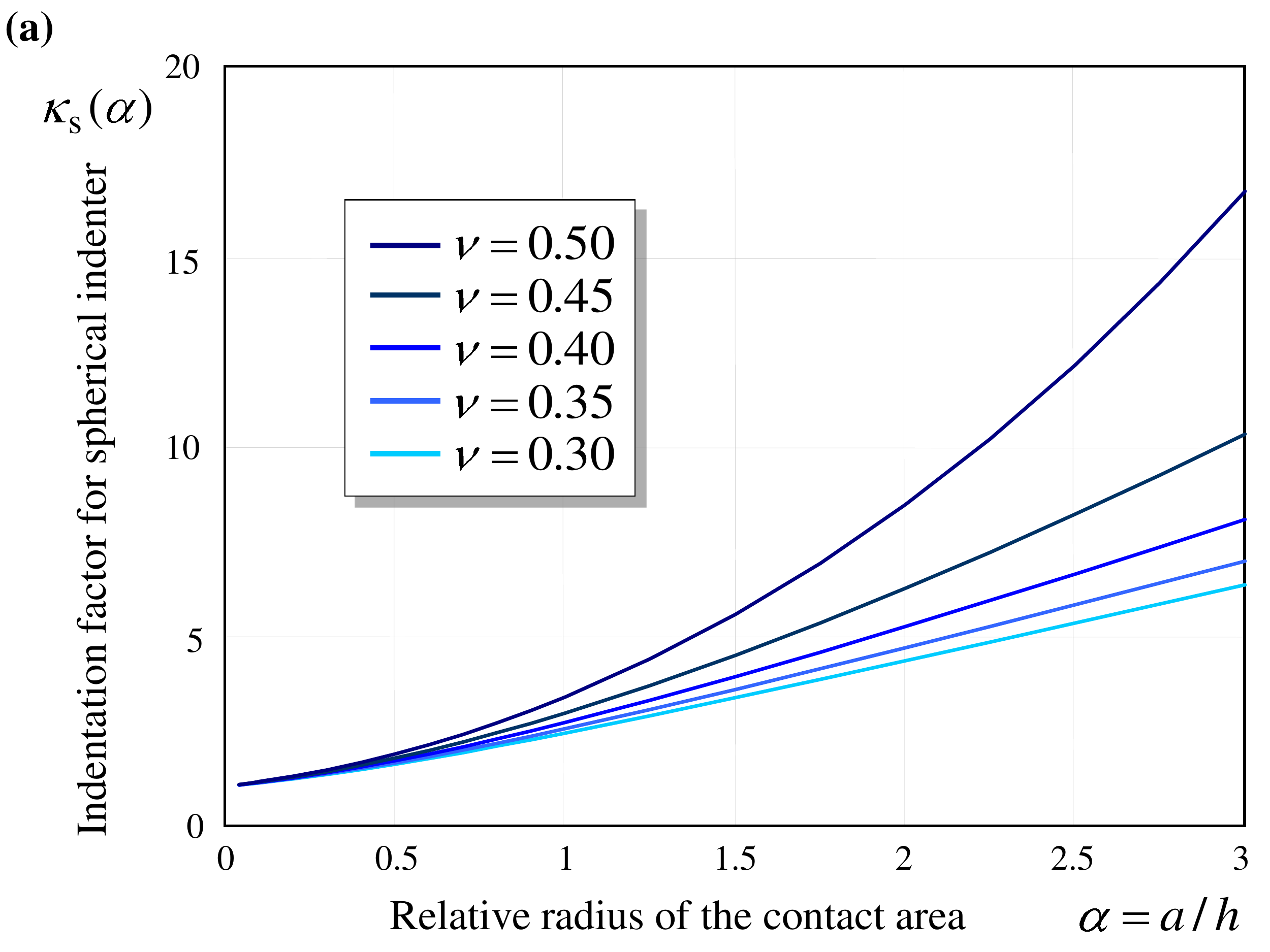}%\hskip-0.3cm
    \includegraphics[scale=0.32]{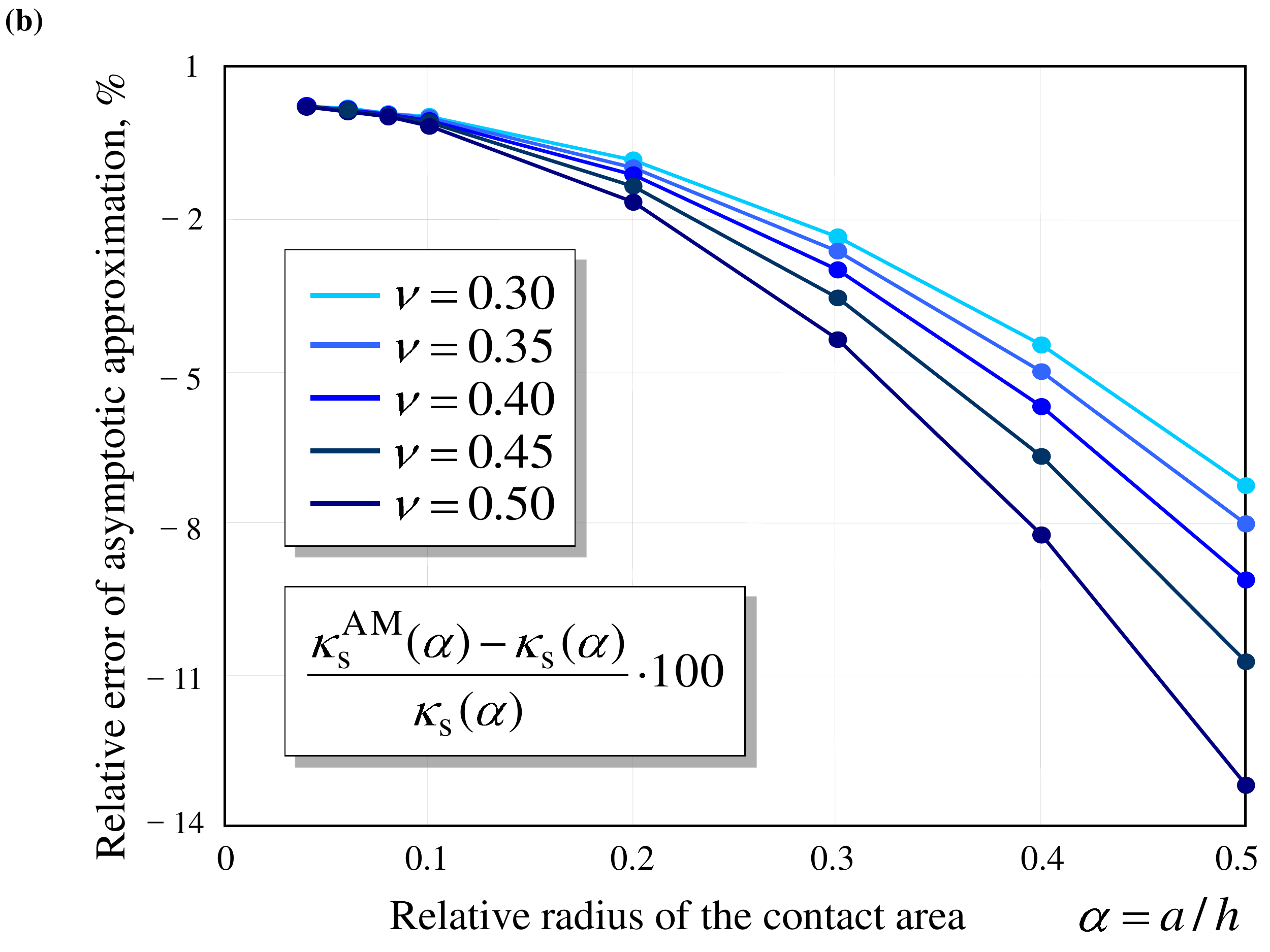}
    }
%\vskip-3.5cm    
    \caption{(a) Indentation scaling factor for the spherical indenter as a function of relative contact radius; 
    (b) Relative error of of the asymptotic approximation (\ref{1Ds(2.20)}).    }
%\vskip-1.0cm        
    \label{Fig-kappa-s}
\end{figure}

Now, employing the forth-order asymptotic model constructed by \citet{Vorovich_et_al_1974,Argatov2002}, we obtain 
\begin{eqnarray}
\kappa_{\rm s}(\alpha) & = & 1+\alpha\frac{2a_0}{\pi}+
\alpha^2\frac{2a_0^2}{\pi^2}
-\alpha^3\biggl[\frac{16a_0^3}{27\pi^3}-\frac{32 a_1}{15\pi}\biggr]
\nonumber \\
{} & {} & {}-\alpha^4\biggl[\frac{22a_0^4}{3\pi^4}+\frac{32 a_0 a_1}{15\pi^2}\biggr]
+O(\alpha^5).
\label{1Ds(2.20)}
\end{eqnarray}

Fig.\,\ref{Fig-kappa-s}a shows details of the behavior of $\kappa_{\rm s}(\alpha)$ for different values of Poisson'a ratio. The errors of the asymptotic approximation $\kappa_{\rm s}^{\rm AM}(\alpha)$ given by (\ref{1Ds(2.20)}) are plotted as functions of $a/h$ in Fig.\,\ref{Fig-kappa-s}b based on the data given in Table~\ref{table:kappa-s}. 

\subsection{Modified incomplete storage modulus and loss angle}
\label{1DsSection2.4}

In the viscoelastic case, according to Eqs.~(\ref{1Ds(2.2)}), (\ref{1Ds(2.3)}), (\ref{1Ds(2.14)}), and (\ref{1Ds(2.15)}), we will have
\begin{equation}
P(t)=\frac{4E_\infty h^3}{3(1-\nu^2)R}
\int\limits_{0}^t\frac{d}{d\tau}\bigl\{
\varpi(\tau)^{3/2}f(\varpi(\tau))\bigr\}\Psi(t-\tau)\,d\tau.
\label{1Ds(2.21)}
\end{equation}
Here we used the notation (cf. (\ref{1Ds(2.14)}))
\begin{equation}
\varpi(t)=\frac{\sqrt{w(t)R}}{h}.
\label{1Ds(2.22)}
\end{equation}

Equation (\ref{1Ds(2.21)}) can be simplified for the case of a viscoelastic half-space when
$f(\varpi(\tau))\equiv 1$ as follows:
\begin{equation}
P(t)=\frac{4\sqrt{R}}{3(1-\nu^2)}\int\limits_{0}^t\frac{d}{d\tau}\bigl\{
w(\tau)^{3/2}\bigr\}E(t-\tau)\,d\tau.
\label{1Ds(2.23)}
\end{equation}

We emphasize that Eqs.~(\ref{1Ds(2.21)}) and (\ref{1Ds(2.23)}) are valid under the assumption that the indenter's displacement $w(t)$ increases in the time interval $(0,t_m)$.

Further, we consider again the same single indentation test with a prescribed sinusoidal displacement according to the indentation protocol (\ref{1Ds(4.1)}). We may use Eqs.~(\ref{1Ds(2.21)}) and (\ref{1Ds(2.23)}) in the time interval $(0,t_m)$ with $t_m=\pi/(2\omega)$, that is up to the moment, when the indenter reaches its maximum indentation depth $w_0$.

By analogy with Eq.~(\ref{1Ds(4.3)}), we consider the quantity 
\begin{equation}
\frac{3(1-\nu^2)}{4\sqrt{R}}\frac{P(t_m)}{w_0^{3/2}}=\int\limits_0^{t_m}
E(t-\tau)\frac{d}{d\tau}\Bigl(
\frac{w(\tau)}{w_0}\Bigr)^{3/2}d\tau.
\label{1Ds(33.1)}
\end{equation}
Note that the quantity on the right-hand side of Eq.~(\ref{1Ds(33.1)}) was previously considered in a number of studies on indentation of viscoelastic materials \citep{Hu_et_al2001,KrenNaumov2010}.

Substituting the expression (\ref{1Ds(4.1)}) into the right-hand side of Eq.~(\ref{1Ds(33.1)}),
we arrive at the following integral with $\beta=3/2$:
\begin{equation}
\tilde{E}_\beta(\omega)=\int\limits_0^{\pi/(2\omega)}
E\Bigl(\frac{\pi}{2\omega}-\tau\Bigr)\frac{d}{d\tau}(\sin\omega\tau)^\beta d\tau.
\label{1Ds(33.2)}
\end{equation}
Observe that here the parameter $\beta$ was introduced to simplify notation. However, later we show (see Remark~\ref{rem2}) that the notation $\tilde{E}_\beta(\omega)$ is meaningful for different values of $\beta$.

By changing the integration variable, the integral (\ref{1Ds(33.2)}) may be cast in the form
\begin{equation}
\tilde{E}_\beta(\omega)=\omega\int\limits_0^{\pi/(2\omega)}
E(s)\beta\cos^{\beta-1}\omega s\sin\omega s\, ds.
\label{1Ds(33.3)}
\end{equation}

It is clear that for $\beta=1$, the right-hand sides of (\ref{1Ds(4.4)}) and (\ref{1Ds(33.3)}) coincide. 
The quantity $\tilde{E}_\beta(\omega)$ will be called the modified incomplete storage modulus.

{\rem{
\label{rem2}
Recall \citep{Galin1946,BorodichKeer2004} that the force-displacement relationship in the elastic case for a rigid blunt indenter with the shape function $z=A r^\lambda$ is given by the equation
$P=[E/(1-\nu^2)]A^{1-\beta} K_\beta w^\beta$ with $\beta=(\lambda+1)/\lambda$ and (with $\Gamma(x)$ being the Gamma function)
$$
K_\beta=\frac{2^{2(\beta-1)}(\beta-1)}{\displaystyle
\beta\exp\Bigl(\frac{1+\beta-\beta^2}{(\beta-1)^2}\ln(\beta-1)\Bigr)}
\Gamma\Bigl(\frac{1}{2(\beta-1)}\Bigr)^{2(1-\beta)}
\Gamma\Bigl(\frac{1}{\beta-1}\Bigr)^{\beta-1}.
$$
For a spherical indenter, we have $\lambda=2$ and $\beta=3/2$. We refer to \citep{Argatov2011} for complete details of this consideration in the elastic case. In the viscoelastic case, the force-displacement relationship in the loading phase is given by
$$
P(t)=\frac{E A^{1-\beta} K_\beta}{1-\nu^2}
\int\limits_{0}^t\frac{d}{d\tau}\bigl\{w(\tau)^\beta\bigr\}E(t-\tau)\,d\tau.
$$
Comparing this equation with Eq.~{\rm (\ref{1Ds(2.23)})}, we see that the blunt indentation test yields the modified incomplete storage modulus $\tilde{E}_\beta(\omega)$ introduced by formula {\rm (\ref{1Ds(33.2)})}. This explains the introduced notation. 
}}

Further, assuming the variation of the indenter displacement in the form (\ref{1Ds(4.1)}), we get the following variation of the contact force:
\begin{equation}
P(t) = \frac{4\sqrt{R}}{3(1-\nu^2)}w_0^{3/2}\int\limits_0^t
E(t-\tau)\frac{d}{d\tau}(\sin\omega \tau)^{3/2}\,d\tau.
\label{1Ds(33.4x)}
\end{equation}
Now, replacing $3/2$ with $\beta$ in the exponent under the integral sign in (\ref{1Ds(33.4x)}), we obtain
\begin{equation}
P(t) = \frac{4\sqrt{R}}{3(1-\nu^2)}w_0^{3/2}\beta\omega\int\limits_0^t
E(s)(\sin\omega(t-s))^{\beta-1}\cos\omega(t-s)\,ds.
\label{1Ds(33.4)}
\end{equation}

Now, let $\tilde{t}_M$ be the time moment when the contact force (\ref{1Ds(33.4)}) reaches its maximum, i.\,e., $\dot{P}(\tilde{t}_M)=0$. Then, by analogy with the case of linear harmonic vibrations, we put
\begin{equation}
\tilde{\delta}_\beta(\omega)=\frac{\pi}{2}-\omega \tilde{t}_M.
\label{1Ds(33.5)}
\end{equation}
The quantity $\tilde{\delta}_\beta(\omega)$ will be called the modified incomplete loss angle determined from the sinusoidally-driven displacement-controlled spherical indentation test. 
In view of (\ref{1Ds(4.1a)}), formula (\ref{1Ds(33.5)}) can be rewritten in the form (\ref{1Ds(4.7)}).

\subsection{Modified incomplete storage modulus and loss angle. Standard viscoelastic solid model}
\label{1DsSection2.5}

In order to fix our ideas, we assume that the layer's material follows a standard linear viscoelastic solid model, which is described by the following normalized creep and relaxation functions:
\begin{equation}
\Phi(t)=1-(1-\rho)\exp(-t/\tau_s),\quad
\Psi(t)=1-(1-1/\rho)\exp(-t/(\rho \tau_s)).
\label{1Ds(ssm.1)}
\end{equation}
Here, $\tau_s$ is the characteristic retardation or creep time of strain under applied step of stress, $\rho$ is the ratio of $E_\infty$ to the unrelaxed elastic modulus $E_0$ (modulus $E(t)$ at $t=0$), i.\,e., $\rho=E_\infty/E_0<1$.

The following relations are well known \citep{Tschoegl1997}:
\begin{equation}
E_1(\omega)=E_\infty+(E_0-E_\infty)\frac{\omega^2(\rho\tau_s)^2}{
\omega^2(\rho\tau_s)^2+1},
\label{1Ds(ssm.2)}
\end{equation}
%\begin{equation}
$$
E_2(\omega)=(E_0-E_\infty)\frac{\omega\rho\tau_s}{
\omega^2(\rho\tau_s)^2+1},
$$
%\label{1Ds(ssm.3)}
%\end{equation}
\begin{equation}
\delta(\omega)=\arctan\frac{(1-\rho)\omega\rho\tau_s}{
\rho+\omega^2(\rho\tau_s)^2}.
\label{1Ds(ssm.4)}
\end{equation}

\begin{figure}[h!]
%\vskip-1.0cm    
    \centering
    \hbox{
    \includegraphics[scale=0.32]{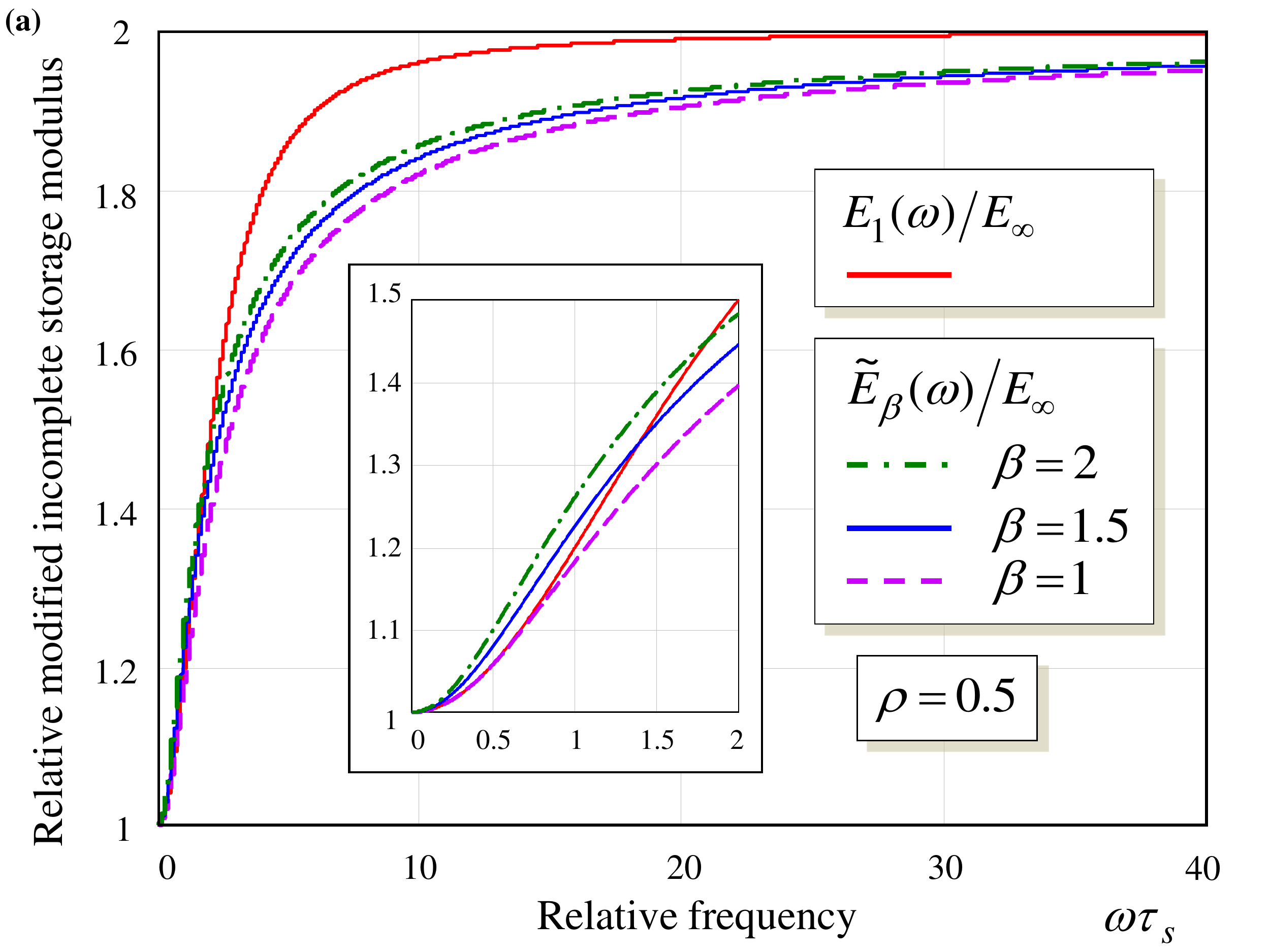}%\hskip-0.3cm
    \includegraphics[scale=0.32]{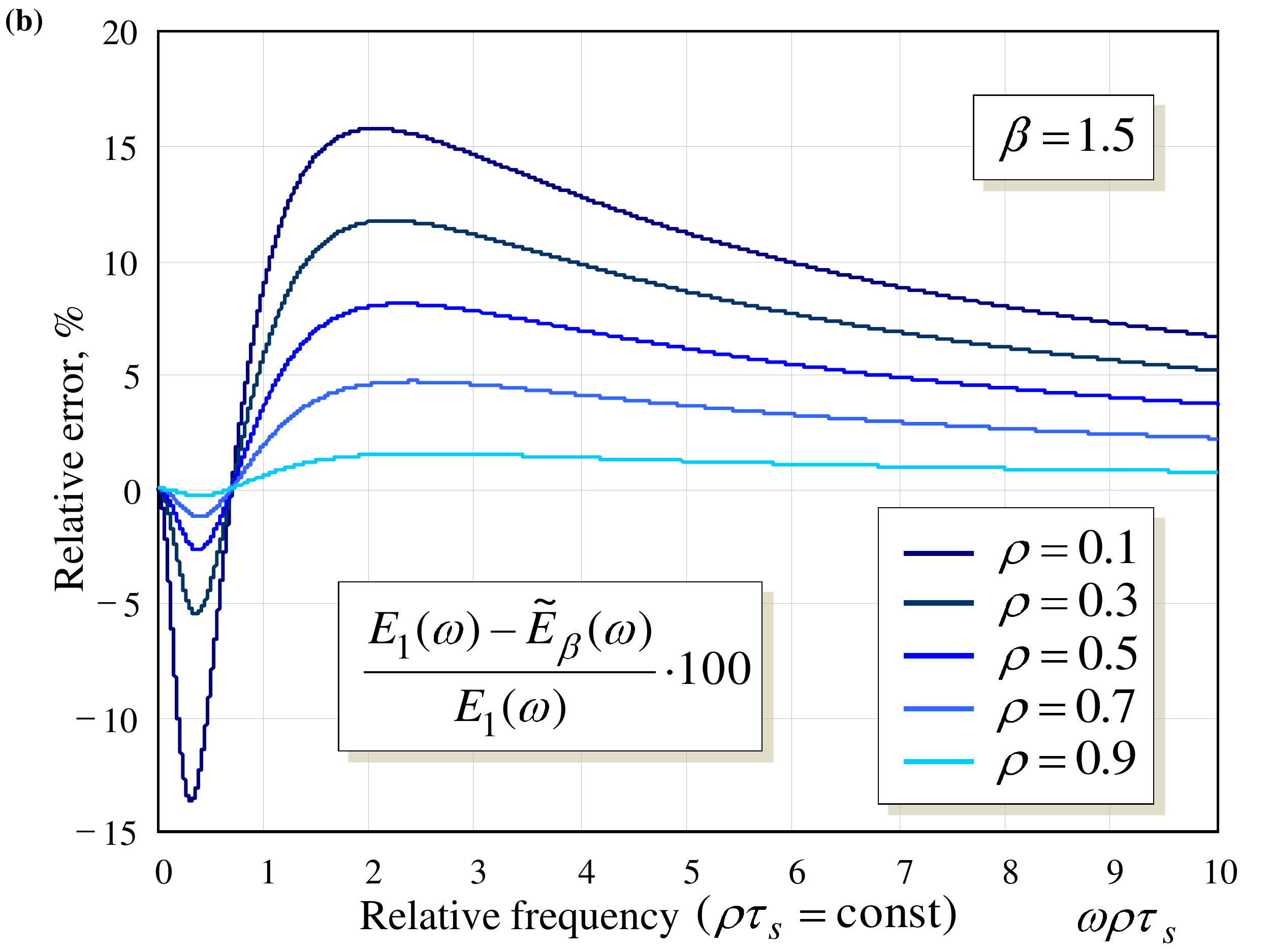}
    }
%\vskip-3.5cm    
    \caption{Modified incomplete storage modulus $\tilde{E}_\beta(\omega)$.    }
%\vskip-1.0cm        
    \label{E_beta-1}
\end{figure}

Fig.~\ref{E_beta-1}a shows the behavior of the modified incomplete storage modulus $\tilde{E}_\beta(\omega)$ in comparison with that of the storage modulus $E_1(\omega)$ given by (\ref{1Ds(ssm.2)}) for $\beta=1$, $1{.}5$, and 2. At high frequencies, the both dimensionless quantities $E_1(\omega)/E_\infty$ and $\tilde{E}_\beta(\omega)/E_\infty$ approach the limit value $E_0/E_\infty=1/\rho=2$.
Fig.~\ref{E_beta-1}b shows the relative error of the approximation of $E_1(\omega)$ by $\tilde{E}_{3/2}(\omega)$ for different values of the dimensionless parameter $\rho$. In each case, it is assumed that the mean relaxation time $\rho\tau_s$ is the same. 

\begin{figure}[h!]
%\vskip-1.0cm    
    \centering
    \hbox{
    \includegraphics[scale=0.32]{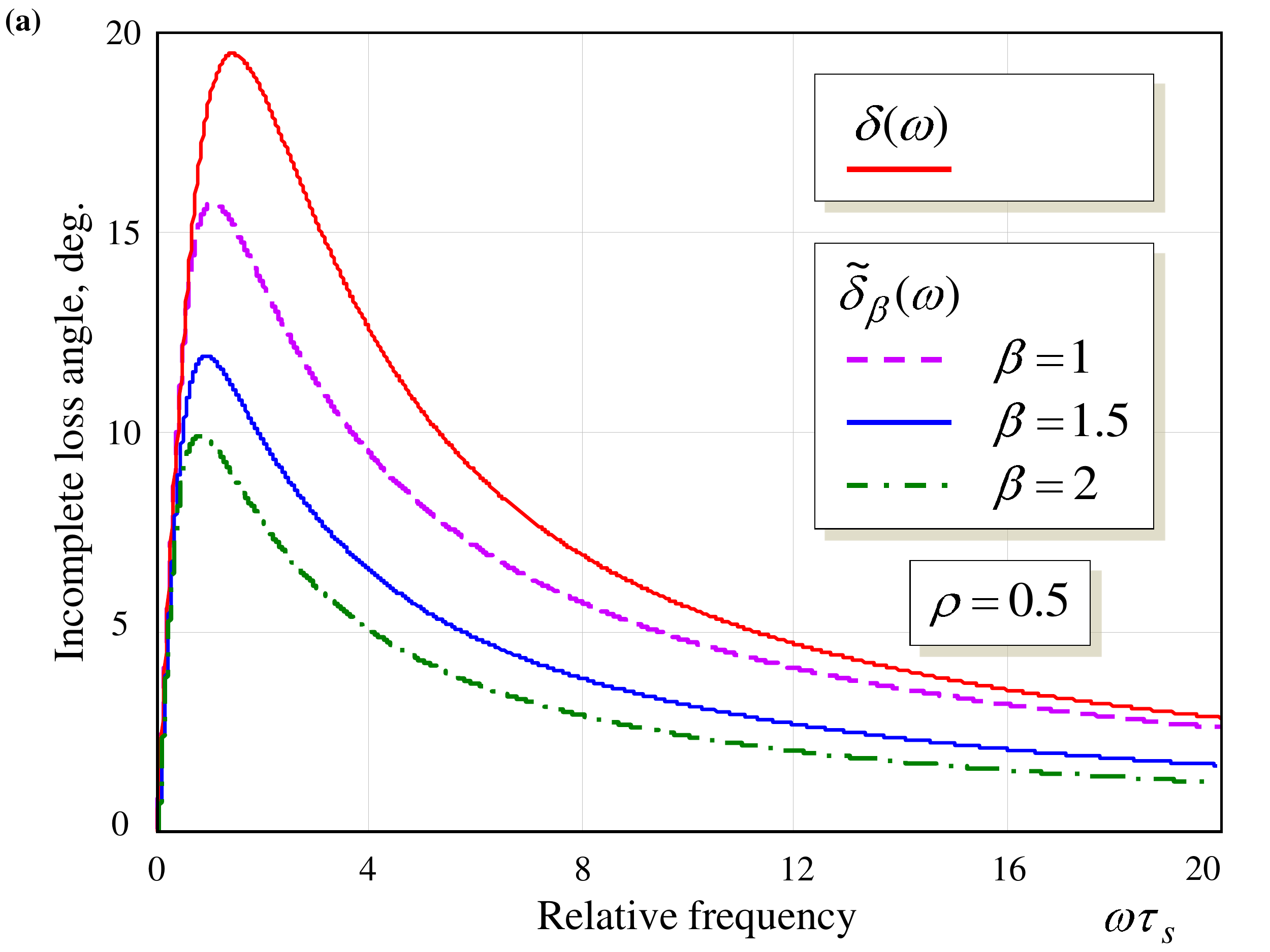}%\hskip-0.3cm
    \includegraphics[scale=0.32]{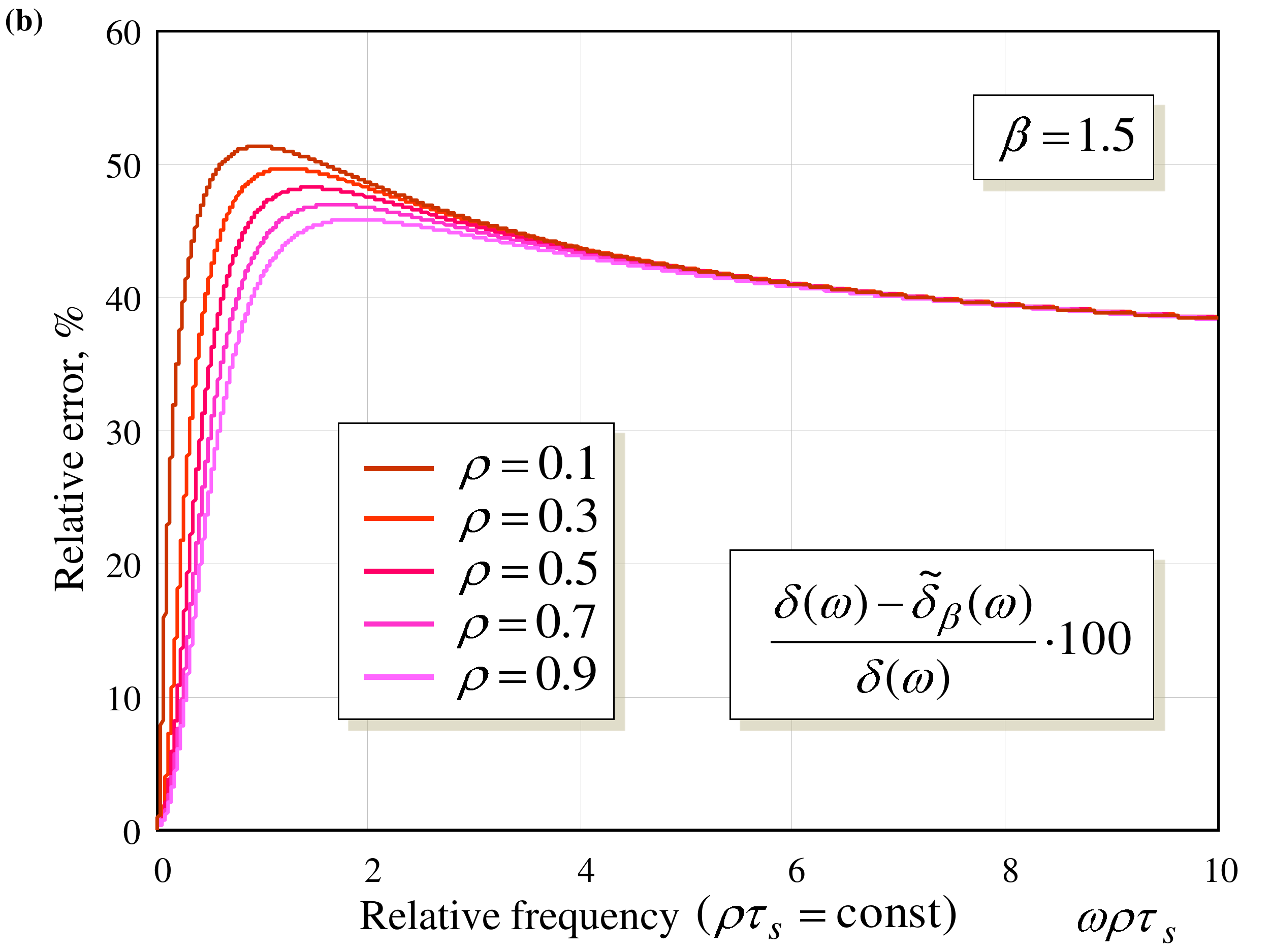}
    }
%\vskip-3.5cm    
    \caption{Modified incomplete loss angle $\tilde{\delta}_\beta(\omega)$.    }
%\vskip-1.0cm        
    \label{E_beta-2}
\end{figure}

Fig.~\ref{E_beta-2}a presents the behavior of the modified incomplete loss angle $\tilde\delta_M(\omega)$ determined from the displacement-controlled indentation test in comparison with that of the loss angle $\delta(\omega)$ given by (\ref{1Ds(ssm.4)}) for $\beta=1$, $1{.}5$, and 2.
The error of the approximation of the loss angle $\delta(\omega)$ by $\tilde\delta_M(\omega)$ is shown in Fig.~\ref{E_beta-2}b. Unfortunately, the deference between $\tilde\delta_M(\omega)$ and $\delta(\omega)$ does not vanish as $\omega\to\infty$. It will be shown that the limit value of the relative error is $33{.}33\%$.

\section{Accounting for the thickness effect in spherical indentation of a viscoelastic layer}
\label{1DsSection3}

\subsection{Dynamic parameters for assessing the mechanical properties and viability of articular cartilage by a spherical indentation test
}
\label{1DsSection3.1}

Finally, let us consider the quantity 
\begin{equation}
\frac{3(1-\nu^2)}{4\sqrt{R}}\frac{P(t_m)}{w_0^{3/2}\kappa_{\rm s}(\alpha_m)}=\tilde{E}_{3/2}^0(\omega,\varpi_0),
\label{1Ds(3.1)}
\end{equation}
where $\kappa_{\rm s}(\alpha_m)$ is the indentation scaling factor corresponding to the maximum indentation depth 
$a_m=a(t_m)$, thus $\alpha_m=a_m/h$.

According to Eqs.~(\ref{1Ds(2.18)}) and (\ref{1Ds(2.21)}), the right-hand side of Eq.~(\ref{1Ds(3.1)}) is determined as follows:
\begin{equation}
\tilde{E}_{3/2}^0(\omega,\varpi_0)=\frac{1}{f(\varpi_0)}\int\limits_0^{\pi/(2\omega)}
E\Bigl(\frac{\pi}{2\omega}-\tau\Bigr)\frac{d}{d\tau}\bigl\{
(\sin\omega\tau)^{3/2}f(\varpi_0\sqrt{\sin\omega\tau})
\bigr\}\,d\tau.
\label{1Ds(3.2)}
\end{equation}
Here we introduced the notation (see Eq.~(\ref{1Ds(2.22)}))
\begin{equation}
\varpi_0=\frac{\sqrt{w_0 R}}{h}.
\label{1Ds(3.3)}
\end{equation}

Now, let $\tilde{t}_M^0$ be the time moment when the contact force (\ref{1Ds(2.21)}) reaches its maximum, i.\,e., $\dot{P}(\tilde{t}_M^0)=0$. Then, by analogy with the modified incomplete loss angle
$\tilde{\delta}_{3/2}(\omega)$, we put
\begin{equation}
\tilde{\delta}_{3/2}^0(\omega)=\frac{\pi}{2}-\omega \tilde{t}_M^0.
\label{1Ds(3.4)}
\end{equation}
In view of (\ref{1Ds(4.1a)}), formula (\ref{1Ds(3.4)}) can be rewritten in the form (\ref{1Ds(4.7)}).

The quantities $\tilde{E}_{3/2}^0(\omega,\varpi_0)$ and $\tilde{\delta}_{3/2}^0(\omega)$ (with explicit dependence on $\varpi_0$ hidden) will be simply called the modified storage modulus and modified loss angle.

It should be emphasized that the quantities $\tilde{E}_{3/2}^0(\omega,\varpi_0)$ and $\tilde{\delta}_{3/2}^0(\omega)$ depend on the layer thickness, although this fact is not reflected in the notation. Thus, in the viscoelastic case, the application of the indentation scaling factor $\kappa_{\rm s}(\alpha_m)$ in the same way as on the right-hand side of formula (\ref{1Ds(3.1)}) does not completely accounts for the thickness effect. 

\begin{figure}[h!]
%\vskip-1.0cm    
    \centering
    \hbox{
    \includegraphics[scale=0.32]{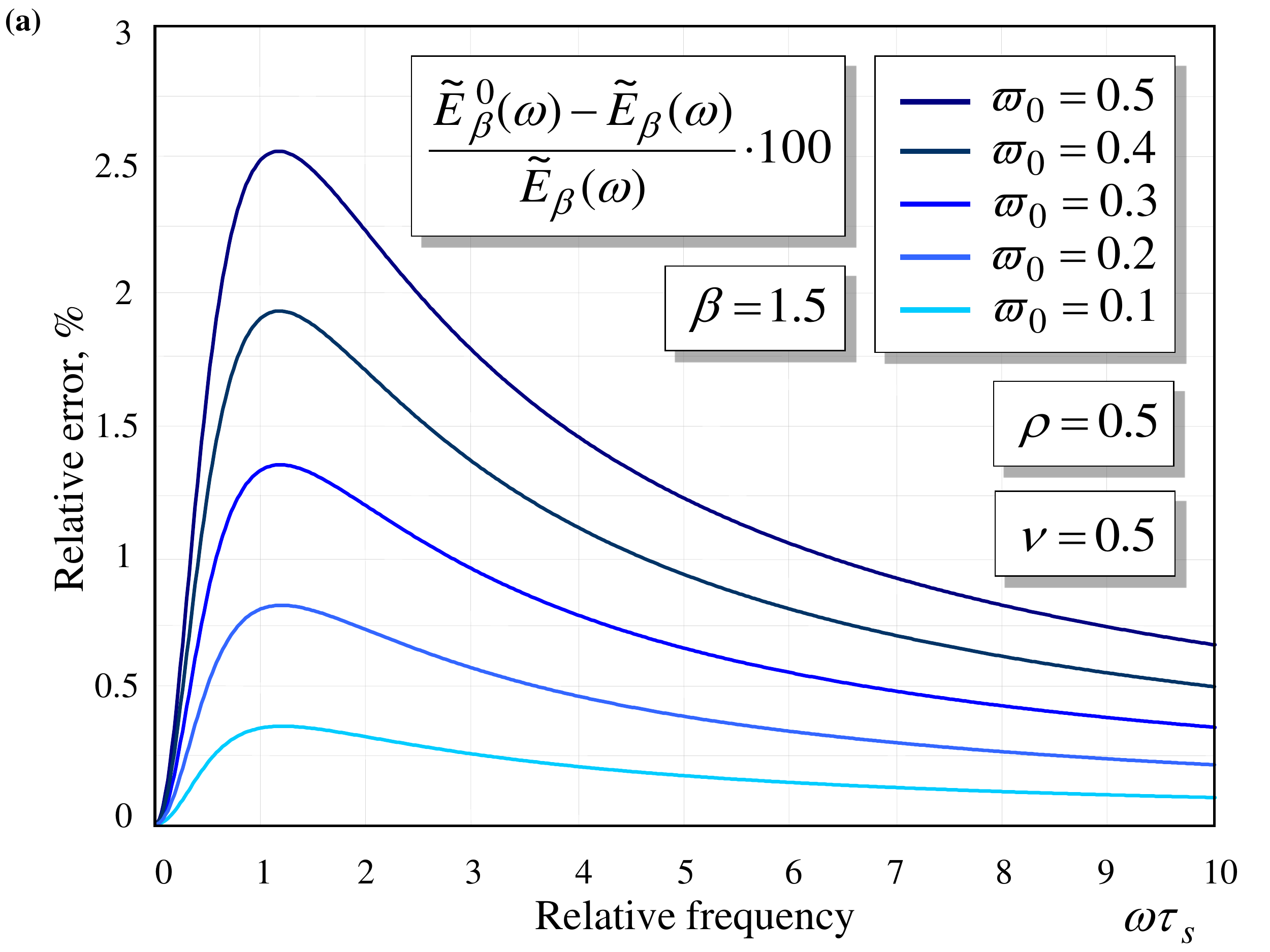}%\hskip-0.3cm
    \includegraphics[scale=0.32]{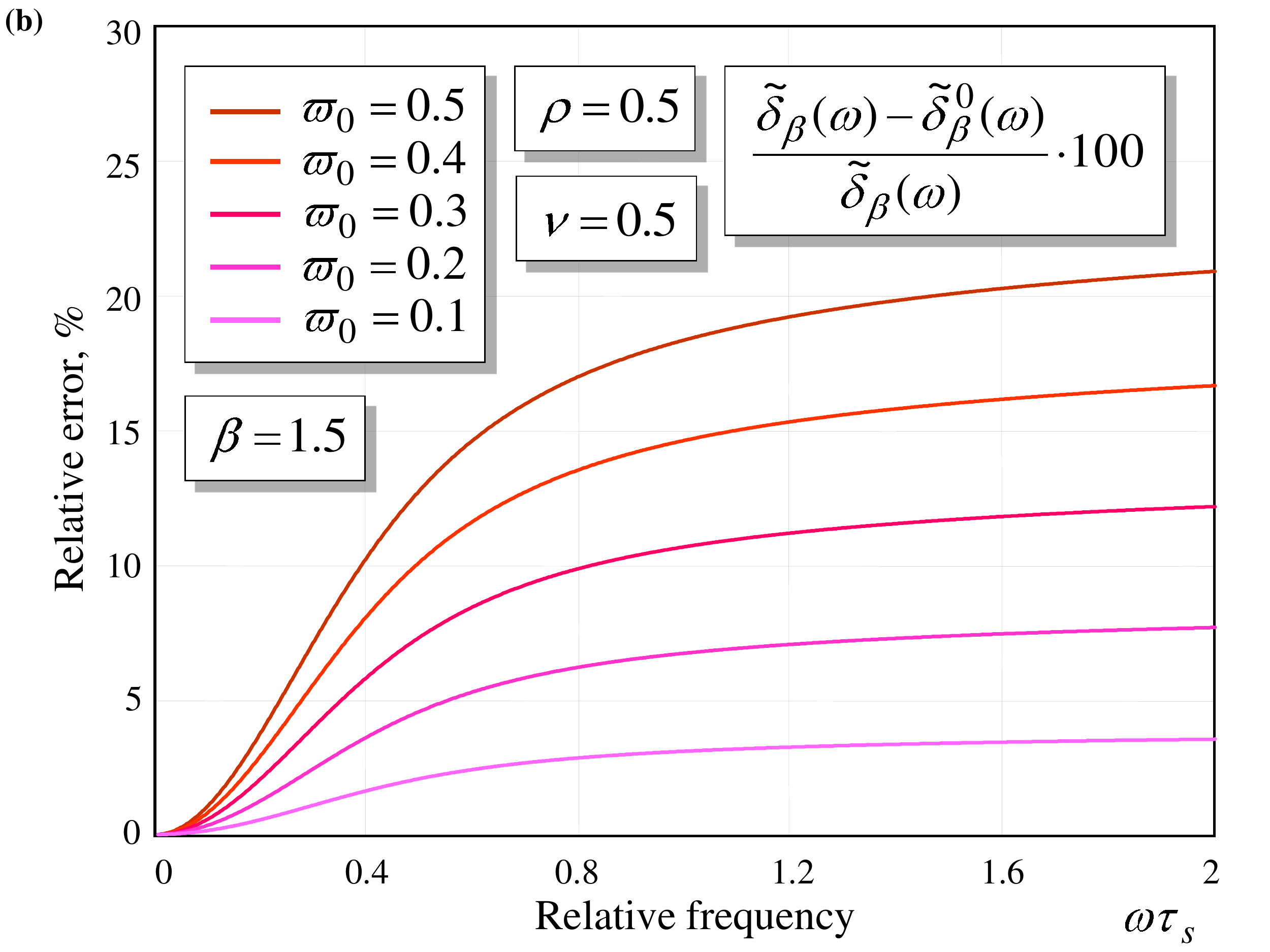} 
    }
%\vskip-3.5cm    
    \caption{Difference between $\tilde{E}_{3/2}^0(\omega,\varpi_0)$, $\tilde{\delta}_{3/2}^0(\omega)$ and $\tilde{E}_{3/2}(\omega)$, $\tilde{\delta}_{3/2}(\omega)$ for different levels of indentation.    }
%\vskip-1.0cm        
    \label{omega_w0n05}
\end{figure}

Fig.~\ref{omega_w0n05}a presents the comparison of the quantity $\tilde{E}_{3/2}^0(\omega,\varpi_0)$ with the modified incomplete storage modulus $\tilde{E}_{3/2}(\omega)$ in the case of standard solid model. As it could be expected, the difference tends to zero as $\omega\to 0$, that is as the indentation protocol approaches the quasi-static limit. In the high-frequency range (as $\omega\to\infty$), it can be also established rigorously that 
$\tilde{E}_{3/2}^0(\omega,\varpi_0)$ and $\tilde{E}_{3/2}(\omega)$ both tend to $E(0)$ as well as $E_1(\omega)$ does. 
On the contrary, the deference between $\tilde\delta_M^0(\omega)$ and $\tilde\delta_M(\omega)$ does not vanish as $\omega\to\infty$ (see Fig.~\ref{omega_w0n05}b), and the limit value of the relative error depends on the value of $\varpi_0$.

\subsection{Asymptotic analysis of $\tilde{E}_{3/2}^0(\omega,\varpi_0)$ in the low- and high-frequency limits}
\label{1DsSection3.2}

To fix our ideas, let us assume that the relaxation modulus $E(t)$ is determined by the Prony series
\begin{equation}
E(t)=E_\infty+\sum_{j=1}^m E_j\exp\Bigl(-\frac{t}{\rho_j}\Bigr),
\label{1Ds(7.1)}
\end{equation}
where $E_j$ and $\rho_j$ are positive constants representing the relaxation strengths and relaxation times. Without loss of generality we may assume that $\rho_1<\rho_2<\ldots<\rho_m$.

Integrating by parts in (\ref{1Ds(3.2)}) and changing the integration variable, we rewrite Eq.~(\ref{1Ds(3.2)}) as
\begin{equation}
\tilde{E}_\beta^0(\omega,\varpi_0)=E(0)+\frac{1}{\omega f(\varpi_0)}\int\limits_0^{\pi/2}
E^\prime\Bigl(\frac{z}{\omega}\Bigr)
(\cos z)^\beta f(\varpi_0\sqrt{\cos z})\,dz.
\label{1Ds(7.2)}
\end{equation}
Here again it is assumed that $\beta=3/2$.

In the low-frequency limit, the behavior of $\tilde{E}_\beta^0(\omega,\varpi_0)$ as $\omega\to 0$ will depend on the asymptotic behavior of the integral 
\begin{equation}
I_\beta^j(\omega,\varpi_0)=\int\limits_0^{\pi/2}\exp\Bigl(-\frac{z}{\omega\rho_j}\Bigr)
(\cos z)^\beta f(\varpi_0\sqrt{\cos z})\,dz.
\label{1Ds(7.3)}
\end{equation}

Using the formula (\citet{Dwight1961}, formula (567.9))
$$
\int x^n\exp(ax)\,dx=\exp(ax)\sum_{k=0}^n\frac{(-1)^{n-k} n!}{k! a^{n+1-k}}x^k,
$$
it can be easily shown that 
\begin{eqnarray}
\int\limits_0^{\pi/2}\exp\Bigl(-\frac{z}{\omega\rho_j}\Bigr)z^n\,dz & = & n!(\omega\rho_j)^{n+1}-\exp\Bigl(-\frac{\pi}{2\omega\rho_j}\Bigr)
\sum_{k=0}^n\frac{n!(\omega\rho_j)^{n+1-k}}{k!}\Bigl(\frac{\pi}{2}\Bigr)^k
\nonumber \\
{} & = & n!(\omega\rho_j)^{n+1}+O\Bigl((\omega\rho_j)^{n+1}
\exp\Bigl(-\frac{\pi}{2\omega\rho_j}\Bigr)\Bigr), \quad\omega\to 0.
\label{1Ds(7.4)}
\end{eqnarray}

Thus, making use of formula (\ref{1Ds(7.4)}) and the two-term Taylor expansion 
$$
\frac{1}{f(\varpi_0)}(\cos z)^\beta f(\varpi_0\sqrt{\cos z})=
1-z^2\biggl(\frac{\beta}{2}+\frac{f^\prime(\varpi_0)}{f(\varpi_0)}\frac{\varpi_0}{4}
\biggr)+O(z^4),
$$
one can now expand the integral $I_\beta^j(\omega,\varpi_0)$ determined by (\ref{1Ds(7.4)}) in power series with respect to $\omega^2$. 
In such a way, we arrive at the following asymptotic representation:
\begin{equation}
\tilde{E}_\beta^0(\omega,\varpi_0)=E_\infty+\omega^2\sum_{j=1}^m E_j\rho_j^2
\biggl(\beta+\frac{f^\prime(\varpi_0)}{f(\varpi_0)}\frac{\varpi_0}{2}
\biggr)+O(\omega^4).
\label{1Ds(7.5)}
\end{equation}
Here it was taken into account that 
\begin{equation}
E(0)=E_\infty+\sum_{j=1}^m E_j.
\label{1Ds(7.55)}
\end{equation}

On the other hand, the following two-term asymptotic expansions holds true for the storage modulus
\citep{Tschoegl1997}:
\begin{equation}
E_1(\omega)=E_\infty+\omega^2\sum_{j=1}^m E_j\rho_j^2+O(\omega^4),\quad\omega\to 0.
\label{1Ds(7.6)}
\end{equation}

Comparing asymptotic expansions (\ref{1Ds(7.5)}) and (\ref{1Ds(7.6)}), we see that the second terms on their right-hand sides coincide only if $\beta=1$ and $\varpi_0=0$, when $\tilde{E}_\beta^0(\omega,\varpi_0)$ coincides with the incomplete storage modulus $\tilde{E}_1(\omega)$.

In the high-frequency limit, the behavior of $\tilde{E}_\beta^0(\omega,\varpi_0)$ as $\omega\to\infty$ depends on the smoothness properties of the function $E^\prime(t)$ at the point $t=0$. In view of (\ref{1Ds(7.1)}), Eq.~(\ref{1Ds(7.2)}) readily yields 
\begin{equation}
\tilde{E}_\beta^0(\omega,\varpi_0)=E(0)+\frac{E^\prime(0)}{\omega}\frac{I_\beta^0(\varpi_0)}{f(\varpi_0)}
+O(\omega^{-2}),
\label{1Ds(7.7)}
\end{equation}
where $E^\prime(0)=-\sum_{j=1}^m E_j/\rho_j$ and
$$
I_\beta^0(\varpi_0)=\int\limits_0^{\pi/2}
(\cos z)^\beta f(\varpi_0\sqrt{\cos z})\,dz.
$$

On the other hand, the following two-term asymptotic expansions holds true for the storage modulus
\citep{Argatov2012}:
\begin{equation}
E_1(\omega)=E(0)-\frac{E^{\prime\prime}(0)}{\omega^2}+O(\omega^{-4}),\quad\omega\to\infty.
\label{1Ds(7.8)}
\end{equation}

Thus, based on the asymptotic expansions (\ref{1Ds(7.7)}) and (\ref{1Ds(7.8)}), it is established that 
$\tilde{E}_\beta^0(\omega,\varpi_0)$ tends to $E(0)$ as $\omega\to\infty$ as well as $E_1(\omega)$ does. 

\subsection{Asymptotic analysis of $\tilde{\delta}_\beta^0(\omega)$ in the low- and high-frequency limits}
\label{1DsSection3.3}

Differentiating both sides of Eq.~(\ref{1Ds(2.21)}) with respect to time and taking into account the sinusoidal protocol (\ref{1Ds(4.1)}), we get
\begin{eqnarray}
\frac{3(1-\nu^2)R}{4 h^3\varpi_0^3}\dot{P}(t) & = &
E(0)\frac{d}{dt}\bigl((\sin\omega t)^\beta f(\varpi_0\sqrt{\sin\omega t})\bigr)
\nonumber \\
{} & {} & {}+\int\limits_{0}^t\frac{d}{d\tau}\bigl\{
(\sin\omega\tau)^\beta f(\varpi_0\sqrt{\sin\omega\tau})\bigr\}E^\prime(t-\tau)\,d\tau.
\label{1Ds(8.1)}
\end{eqnarray}

Substituting now the value (see Eq.~(\ref{1Ds(3.4)}))
$$
\tilde{t}_M^0=\frac{1}{\omega}\Bigl(\frac{\pi}{2}-\tilde{\delta}_\beta^0\Bigr)
$$
into the equation $\dot{P}(\tilde{t}_M^0)=0$ in view of (\ref{1Ds(8.1)}), we arrive at the following equation:
\begin{eqnarray}
E(0)\omega \sin\tilde{\delta}_\beta^0\bigl(\cos\tilde{\delta}_\beta^0\bigr)^{\beta-1}\mathcal{F}_\beta^0(\tilde{\delta}_\beta^0)
& = & \int\limits_{0}^{\pi/2-\tilde{\delta}_\beta^0}
\frac{d}{dz}\Bigl\{
\bigl(\cos(\tilde{\delta}_\beta^0+z)\bigr)^\beta 
\nonumber \\
{} & {} & {}\times
f\bigl(\varpi_0\sqrt{\cos(\tilde{\delta}_\beta^0+z)}\bigr)\Bigr\}E^\prime\Bigl(\frac{z}{\omega}\Bigr)\,dz.
\label{1Ds(8.2)}
\end{eqnarray}
Here we introduced the notation
\begin{equation}
\mathcal{F}_\beta^0(\tilde{\delta}_\beta^0)=\beta f\bigl(\varpi_0\sqrt{\cos\tilde{\delta}_\beta^0}\bigr)
+\frac{\varpi_0}{2}\sqrt{\cos\tilde{\delta}_\beta^0}
f^\prime\bigl(\varpi_0\sqrt{\cos\tilde{\delta}_\beta^0}\bigr).
\label{1Ds(8.4)}
\end{equation}

Let $\mathcal{L}(\tilde{\delta}_\beta^0,\omega)$ and $\mathcal{R}(\tilde{\delta}_\beta^0,\omega)$ denote the left and right hand sides of Eq.~(\ref{1Ds(8.2)}), respectively. Following \citet{Argatov2012}, we construct solutions to Eq.~(\ref{1Ds(8.2)}), assuming that $\tilde\delta_M^0(\omega)\simeq C_0\omega$ as $\omega\to 0$ and $\tilde\delta_M^0(\omega)\simeq C_\infty/\omega$ as $\omega\to\infty$, where $C_0$ and $C_\infty$ are constants. In both cases, $\tilde\delta_M^0(\omega)\ll 1$ such that 
\begin{equation}
\mathcal{L}(\tilde{\delta}_\beta^0,\omega)=E(0)\omega\tilde{\delta}_\beta^0 
\mathcal{F}_\beta^0(0)+O\bigl((\tilde{\delta}_\beta^0)^3\bigr), \quad \tilde{\delta}_\beta^0\to 0,
\label{1Ds(8.3)}
\end{equation}
where according to (\ref{1Ds(8.4)}) we have
$$
\mathcal{F}_\beta^0(0)=\beta f(\varpi_0)+\frac{\varpi_0}{2}f^\prime(\varpi_0).
$$

In the low-frequency limit, making use of the asymptotic formula (\ref{1Ds(7.4)}), we get
\begin{equation}
\mathcal{R}(\tilde{\delta}_\beta^0,\omega)=\omega\tilde{\delta}_\beta^0 
\mathcal{F}_\beta^0(0)\sum_{j=1}^m E_j+\omega^2\mathcal{F}_\beta^0(0)\sum_{j=1}^m E_j\rho_j
+O(\omega^3), \quad \omega\to 0.
\label{1Ds(8.5)}
\end{equation}

From (\ref{1Ds(8.3)}) and (\ref{1Ds(8.5)}), it follows that 
\begin{equation}
\tilde{\delta}_\beta^0(\omega) \simeq \frac{\omega}{E_\infty}\sum_{j=1}^m E_j\rho_j
+O(\omega^2), \quad \omega\to 0.
\label{1Ds(8.6)}
\end{equation}
Here the relation (\ref{1Ds(7.55)}) was taken into account. 

We emphasize that the asymptotic representation (\ref{1Ds(8.6)}) is in complete agreement with the leading  term of the asymptotic expansion for the loss angle $\delta(\omega)$ as $\omega\to 0$. 

In the high-frequency limit, we will have
\begin{equation}
\mathcal{R}(\tilde{\delta}_\beta^0,\omega)=
\int\limits_{0}^{\pi/2}\frac{d}{dz}\bigl\{
(\cos z)^\beta f(\varpi_0\sqrt{\cos z})\bigr\}E^\prime(0)\,dz+O(\omega^{-1}),\quad\omega\to\infty.
\label{1Ds(8.7)}
\end{equation}

Now, from (\ref{1Ds(8.3)}) and (\ref{1Ds(8.7)}), it follows that 
\begin{equation}
\tilde{\delta}_\beta^0(\omega) \simeq -\frac{E^\prime(0)}{E(0)\omega}
\frac{f(\varpi_0)}{\mathcal{F}_\beta^0(0)}, \quad \omega\to \infty.
\label{1Ds(8.8)}
\end{equation}

On the other hand, the following asymptotic representation holds true for the loss angle:
\begin{equation}
\delta(\omega) \simeq -\frac{E^\prime(0)}{E(0)\omega}, \quad \omega\to \infty.
\label{1Ds(8.9)}
\end{equation}

Comparing relations (\ref{1Ds(8.8)}) and (\ref{1Ds(8.9)}), we see that they coincide only if $\beta=1$ and 
$f(\varpi_0)\equiv 1$. In the general case, in view of (\ref{1Ds(8.4)}), we have
\begin{equation}
\frac{\delta(\omega)-\tilde{\delta}_\beta^0(\omega)}{\delta(\omega)} \simeq 
\frac{2(\beta-1)f(\varpi_0)+\varpi_0 f^\prime(\varpi_0)}{
2\beta f(\varpi_0)+\varpi_0 f^\prime(\varpi_0)}, \quad \omega\to \infty.
\label{1Ds(8.10)}
\end{equation}

Thus, according to (\ref{1Ds(8.10)}), in the case of a viscoelastic half-space, when $f(\varpi_0)\equiv 1$, the relative error of the approximation $\tilde{\delta}_\beta^0(\omega)$ for $\delta(\omega)$ approaches the value $(\beta-1)/\beta\cdot 100\%$. 

\section{Discussion}
\label{1DsSectionD}

The new material characteristics introduced above, that is the incomplete storage modulus $\tilde{E}_1(\omega)$, the modified incomplete storage modulus $\tilde{E}_\beta(\omega)$, and the modified storage modulus $\tilde{E}_\beta^0(\omega,\varpi_0)$, can be represented as follows: 
\begin{equation}
\tilde{E}_1(\omega) = -\int\limits_{0}^{\pi/(2\omega)} 
E(s)\frac{d}{ds}\{\cos\omega s\}\,ds,
\label{1Ds(d.1)}
\end{equation}
\begin{equation}
\tilde{E}_\beta(\omega)=-\int\limits_0^{\pi/(2\omega)}
E(s)\frac{d}{ds}\bigl\{(\cos\omega s)^\beta\bigr\}\,ds,
\label{1Ds(d.2)}
\end{equation}
\begin{equation}
\tilde{E}_\beta^0(\omega,\varpi_0)=-\int\limits_0^{\pi/(2\omega)}
E(s)\frac{d}{ds}
\Bigl\{(\cos\omega s)^\beta\frac{f(\varpi_0\sqrt{\cos\omega s})}{f(\varpi_0)}\Bigr\}\,ds.
\label{1Ds(d.3)}
\end{equation}

In the same way, the storage modulus $E_1(\omega)$ is recast as 
\begin{equation}
E_1(\omega) = -\int\limits_{0}^\infty E(s)\frac{d}{ds}\{\cos\omega s\}\,ds.
\label{1Ds(d.4)}
\end{equation}

Thus, comparing formulas (\ref{1Ds(d.1)})\,--\,(\ref{1Ds(d.3)}) with (\ref{1Ds(d.4)}), we see that $\tilde{E}_1(\omega)$, $\tilde{E}_\beta(\omega)$, and $\tilde{E}_\beta^0(\omega,\varpi_0)$, 
represent a hierarchy of approximations for $E_1(\omega)$. Applying an asymptotic modeling approach for analyzing the interrelations between the new quantities, we have shown that 
the modified storage moduli asymptotically coincide with the storage modulus in the low- and high-frequency ranges. 

The values $\beta=1$, $\beta=3/2$, and $\beta=2$ correspond respectively to the cases of cylindrical, spherical, and conical indenters. The latter case also applies to pyramidal indenters
\citep{Giannakopoulos2006,Argatov2011}. 

\begin{figure}[h!]
%\vskip-1.0cm    
    \centering
    %\hbox{
    \includegraphics[scale=0.35]{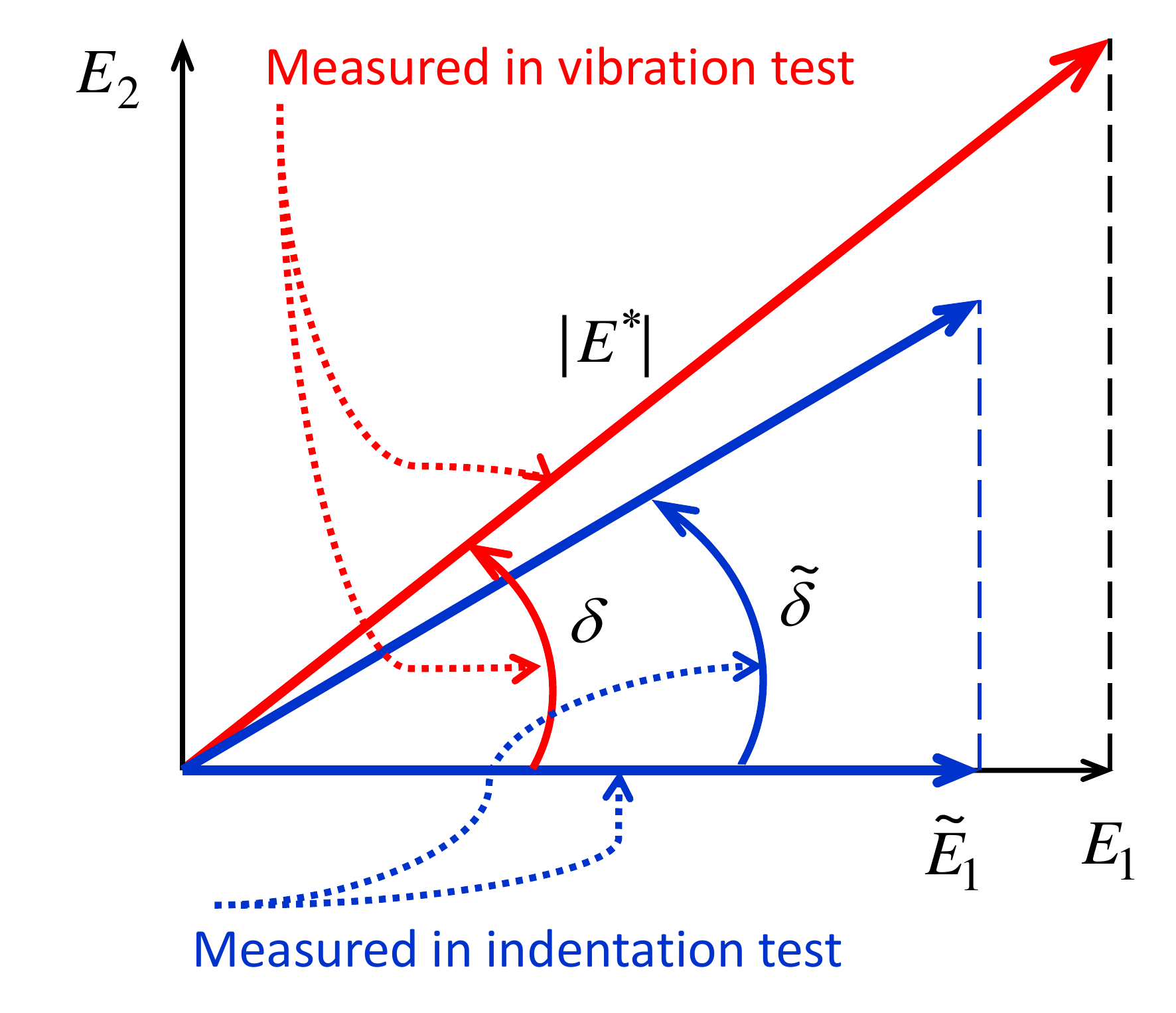}
    %}
%\vskip-3.5cm    
    \caption{Complex dynamic modulus schematic diagram.    }
%\vskip-1.0cm        
    \label{Fig-E1E2}
\end{figure}

Fig.~\ref{Fig-E1E2} illustrates the relationship between the parameters of viscoelastic materials measured in a vibration indentation test and in a single indentation test with a flat-ended cylindrical indenter. 
Due to the nonmonotonic behavior of the modified incomplete storage modulus $\tilde{E}_\beta(\omega)$ with respect to the storage modulus $E_1(\omega)$ as it was shown in Fig.~\ref{E_beta-1} (for the standard viscoelastic solid model), the relationship between the parameters measured in the vibration and indentation tests with a spherical indenter will be more complicated. 

Now, let us consider the application of the developed theory to experimental data \citep{Ronken_et_al_2011}. The experimental setup was described in detail elsewhere \citep{Wirz_et_al_2008}. 
The modified moduli $\tilde{E}_{3/2}^0(\omega,\varpi_0)$ and loss angles $\tilde{\delta}_{3/2}^0(\omega)$ of swine hyaline cartilage at two different locations calculated using the following formulas (see, Eqs.~(\ref{1Ds(3.1)}) and (\ref{1Ds(4.7)})):
\begin{equation}
\tilde{E}_{3/2}^0(\omega,\varpi_0)=\frac{3(1-\nu^2)}{4\sqrt{R}}\frac{P(\tilde{t}_m)}{w_0^{3/2}f(\varpi_0)},
\label{1Ds(33.1x)}
\end{equation}
\begin{equation}
\tilde{\delta}_{3/2}^0(\omega)=\frac{\pi}{2}\frac{(\tilde{t}_m- \tilde{t}_M)}{\tilde{t}_m}.
\label{1Ds(44.7)}
\end{equation}
Here, $f(\varpi_0)=\kappa_{\rm s}(\alpha_m)$ is the indentation scaling factor corresponding to the maximum indentation depth and calculated according to Eq.~(\ref{1Ds(2.19)}). A Poisson's ratio of $0{.}5$ was assumed. Note that the symbol $\tilde{t}_m$ now denotes the time moment of maximum indentation instead of the symbol $t_m$, because in the impact tests the indentation variation $w(t)$ does not follow the sine law (\ref{1Ds(4.1)}) precisely.

\begin{table}[!h]
\caption{Mean and standard deviation of the main parameters of two impact indentation tests (ten repetitions on one spot) for two swine cartilage samples with a spherical impactor of radius $R=0{.}5$~mm and mass $m=1{.}9$~g \citep{Ronken_et_al_2011}.}    
\vskip0.2cm
\begin{center}
\begin{tabular}{l|c|c}\hline
{ } & Lateral condyle & Medial condyle  \\ \hline
Sample thickness, $h$ (mm) &	1.7 &	1.9 \\ 
Initial indenter velocity, $v_0$ (m/s) &	$0{.}249\pm 0{.}002$ &	$0{.}266\pm 0{.}002$ \\ 
Time to maximum contact force, $\tilde{t}_M$ (ms) &	$0{.}68\pm 0{.}01$ &	$0{.}73\pm 0{.}01$ \\ 
Maximum contact force, $P(\tilde{t}_M)$ (N) &	$1{.}04\pm 0{.}02$ &	$0{.}96\pm 0{.}01$ \\
Indentation duration, $\tilde{t}_m$ (ms) &	$0{.}74\pm 0{.}01$ &	$0{.}81\pm 0{.}01$ \\ 
Maximum indentation, $w_0$ (mm) &	$0{.}125\pm 0{.}002$ &	$0{.}145\pm 0{.}003$ \\ 
Contact force at maximum indentation, $P(\tilde{t}_m)$ (N) &	$1{.}02\pm 0{.}02$ &	$0{.}93\pm 0{.}01$ \\ 
\hline
Effective angular frequency, $\omega$ ($\times 10^3\, {\rm rad/s}$) &	$2{.}13\pm 0{.}02$ &	$1{.}93\pm 0{.}02$ \\ 
Level of indentation, $\varpi_0$ &	$0{.}147\pm 0{.}001$ &	$0{.}158\pm 0{.}002$ \\ 
Modified storage modulus, $\tilde{E}_{3/2}^0(\omega,\varpi_0)$ (MPa) &	$15{.}3\pm 0{.}5$ &	$11{.}0\pm 0{.}3$ \\ 
Modified loss angle, $\tilde{\delta}_{3/2}^0(\omega)$ (rad) &	$0{.}114\pm 0{.}007$ &	$0{.}167\pm 0{.}008$ \\ 
Coefficient of restitution, $e_*$ &	$0{.}777\pm 0{.}006$ &	$0{.}722\pm 0{.}012$ \\ 
 \hline
\end{tabular}    
\end{center}
\label{table:test}
\end{table}

The data shown in the upper part of Table~\ref{table:test} was directly assessed in experiments, while the lower part of the table displays results evaluated according to the theory developed herein.  
The example illustrates the fact that the introduced characteristics $\tilde{E}_{3/2}^0(\omega,\varpi_0)$ and $\tilde{\delta}_{3/2}^0(\omega)$ depend on the effective angular frequency $\omega$, which in turn depends on the initial indenter velocity $v_0$ as well as on the mechanical properties of the sample itself. As it could be expected, at high frequencies, the modulus $\tilde{E}_{3/2}^0(\omega,\varpi_0)$ increases with increasing $\omega$, while the angle $\tilde{\delta}_{3/2}^0(\omega)$ decreases (see Table~\ref{table:test}). Note also that this example demonstrates a correlation in behavior of the modified loss angle $\tilde{\delta}_{3/2}^0(\omega)$ and the coefficient of restitution $e_*$. 
Of course, the impact indentation test requires a special consideration, but the observed characteristic behavior of $\tilde{E}_{3/2}^0(\omega,\varpi_0)$ and $\tilde{\delta}_{3/2}^0(\omega)$ with the change in $\omega$ is quite typical.

Now, let us discuss the significance of the developed mathematical approach from the viewpoint of formulating  criteria for evaluation the quality of articular cartilage. In the cylindrical (flat-ended) and spherical dynamic indentation tests, the following cartilage stiffness-related characteristics can be evaluated (see Eqs.~(\ref{1Ds(4.3)}) and (\ref{1Ds(3.1)}), respectively):
\begin{equation}
\frac{1-\nu^2}{2a\kappa_{\rm c}(\alpha)}\frac{P(t_m)}{w_0},
\label{1Ds(4.3C)}
\end{equation}
\begin{equation}
\frac{3(1-\nu^2)}{4\sqrt{R}}\frac{P(t_m)}{w_0^{3/2}\kappa_{\rm s}(\alpha_m)}.
\label{1Ds(3.1C)}
\end{equation}
Here, $w_0=w(t_m)$ is the maximum indentation depth, $P(t_m)$ is the contact force corresponding to the time moment $t=t_m$, when the indenter reaches its maximum indentation depth. The criteria (\ref{1Ds(4.3C)}) and (\ref{1Ds(3.1C)}) in their static form (with no attention paid to the dynamic nature of indentation process) have been used in a number of experimental studies on detection of degenerative changes in joint cartilage.

First of all, it should be noted that the indentation scaling factors $\kappa_{\rm c}(\alpha)$ and $\kappa_{\rm s}(\alpha_m)$ were evaluated under the assumption of isotropy and homogeneity of articular cartilage layer. It is anticipated that the inhomogeneity effect will be smaller in spherical indentation. In view of the layered structure of articular cartilage, the the anisotropy effect requires tacking into account the adjusted value of indentation scaling factor and the corresponding Poisson's ratio. Because Poisson's ratio $\nu$ enters formulas (\ref{1Ds(4.3C)}) and (\ref{1Ds(3.1C)}) not only through the factor $1-\nu^2$ but also through the dependence of $\kappa_{\rm c}(\alpha)$ and $\kappa_{\rm s}(\alpha_m)$ on $\nu$, the question of the appropriate choice of $\nu$ for the criteria (\ref{1Ds(4.3C)}) and (\ref{1Ds(3.1C)}) is more than academic, and it still remains open.

Second, in dynamic indentation testing, the criteria (\ref{1Ds(4.3C)}) and (\ref{1Ds(3.1C)}) will depend on the loading protocol employed, because articular cartilage exhibits viscoelastic and poroelastic properties. Thus, in order to increase the sensitivity of the measurements with a hand-held indentation probe, the indentation protocol should be reproducible as well as the indentation time $t_m$ should be kept the same.

Third, in the framework of linear viscoelasticity, the criteria (\ref{1Ds(4.3C)}) and (\ref{1Ds(3.1C)}) are interpreted as the incomplete storage modulus $\tilde{E}_1(\omega)$ and the modified storage modulus 
$\tilde{E}_{3/2}^0(\omega,\varpi_0)$. Due to the linearity of the theory under consideration, the criterium (\ref{1Ds(4.3C)}) does not depend on the level of indentation. Hence, this fact could be used to check whether the linearity assumption is appropriate for small indentation depths, when $w_0/h\leq 0{.}1$ or even less. On the other hand, the criterium (\ref{1Ds(3.1C)}) does depend on the level of indentation determined by the parameter (see Eq.~(\ref{1Ds(3.3)}))
$$
\varpi_0=\frac{\sqrt{w_0 R}}{h}.
$$
However, since the main manifestation of the thickness effect has been taken into account by means of the indentation scaling factor $\kappa_{\rm s}(\alpha_m)$ evaluated at the maximum indentation depth, the dependence of $\tilde{E}_{3/2}^0(\omega,\varpi_0)$ on $\varpi_0$ is rather weak as it is predicted by the standard viscoelastic solid model (see Fig.~\ref{omega_w0n05}a).

Further, in both indentation tests, one can also measure a dimensionless quantity that is directly related to time-dependent energy dissipation due to viscoelastic and poroelastic relaxation. Namely, the incomplete loss angle $\tilde{\delta}(\omega)$ and the modified loss angle $\tilde{\delta}_{3/2}^0(\omega)$ were introduced based on the time shift between maxima of of the input (indentation displacement) and the output (contact force) through Eq.~(\ref{1Ds(4.7)}). It is important to emphasize that the quantity  $\tilde{\delta}(\omega)$, which is measured in the flat-ended indentation test, does not depend on the thickness of the articular cartilage layer. At the same time, the thickness effect plays a crucial role in manifestation of the time-dependent response to indentation with a spherical indenter (see Fig.~\ref{omega_w0n05}b). 

Finally, the developed viscoelastic models of cylindrical and spherical dynamic indentation tests allow one to compare the diagnostics criteria (that is diagnostics characteristics) experimentally measured by different indentation probes utilizing different indentation protocols (e.g., more closely approximating the actual movement of an operator's hand) as well as operating in different modes (vibration, dynamic indentation, impact testing). In view of the established fact that the incomplete storage modulus $\tilde{E}_1(\omega)$ and the modified storage modulus $\tilde{E}_{3/2}^0(\omega,\varpi_0)$ asymptotically coincide with the storage modulus $E_1(\omega)$ in the low- and high-frequency ranges, the interrelationships between the different tests will be of particular interest in the middle frequency range. 

\section{Conclusions}
\label{1DsSectionC}

We considered frictionless flat-ended and spherical sinusoidally-driven indentation tests utilizing displacement-controlled loading protocol. In order to perform a rigorous analysis, we modeled the deformational behavior of articular cartilage tissue in the framework of viscoelasticity with a time-independent Poisson's ratio. 
In the linear case of flat-ended indentation test, evaluating the dynamic indentation stiffness at the test turning point $t=t_m$, we introduced the incomplete storage modulus 
$\tilde{E}_1(\omega)$ for the effective frequency $\omega=\pi/(2t_m)$. Considering the time difference $t_m-\tilde{t}_M$ between the time moments when the contact force reaches its maximum (dynamic stiffness vanishes at $t=\tilde{t}_M$) and the indenter displacement reaches its maximum (dynamic stiffness becomes infinite at $t=t_m$), we introduced the so-called incomplete loss angle $\tilde{\delta}(\omega)$. 

Analogous quantities were introduced in the nonlinear case of spherical sinusoidally-driven indentation test. First, when the sample thickness effect can be neglected, we introduced the modified incomplete storage modulus $\tilde{E}_{3/2}(\omega)$ and the modified incomplete loss angle $\tilde{\delta}_{3/2}(\omega)$ (we use the same notation as in the linear case). 
Second, to account for the thickness effect, we introduced the indentation scaling factor $\kappa_{\rm s}(\alpha)$ for the spherical indenter depending on Poisson's ratio and the relative contact radius $\alpha=a/h$. Making use of the indentation scaling factor corresponding to the maximum indentation depth $\kappa_{\rm s}(\alpha_m)$, we introduced the modified storage modulus $\tilde{E}_{3/2}^0(\omega,\varpi_0)$ which depends on the level of indentation characterized by the parameter $\varpi_0=\sqrt{w_0 R}/h$. The modified loss angle $\tilde{\delta}_{3/2}^0(\omega)$ was introduced in the same way. 

We applied an asymptotic modeling approach for analyzing the interrelations between the new quantities $\tilde{E}_{3/2}^0(\omega,\varpi_0)$, $\tilde{\delta}_{3/2}^0(\omega)$ and the classical characteristics $E_1(\omega)$, $\delta(\omega)$ in the low- and high-frequency ranges. It was shown that the modified storage modulus asymptotically coincides with the storage modulus in the both limit cases, that is $\tilde{E}_{3/2}^0(\omega,\varpi_0)\simeq E_1(\omega)$ as $\omega\to 0$ and $\omega\to\infty$. However, coinciding with the loss angle $\delta(\omega)$ in the low-frequency range, the modified loss angle $\tilde{\delta}_\beta^0(\omega)$ markedly differs from $\delta(\omega)$ in the high-frequency limit. We illustrated these facts for the standard viscoelastic solid model.

Finally, the present study suggests that the use of dynamic indentation tests is largely twofold: the criteria (\ref{1Ds(4.3C)}) and (\ref{1Ds(3.1C)}) yield dimensional diagnostics characteristics, which could be related to some integral measure of material properties of the tested biological tissue; the second aspect of dynamic indentation diagnostics hinges on the importance of continuous monitoring of the tissue response to indentation. 
It is believed that both characteristics $\tilde{E}_1(\omega)$ (evaluated according to the criterium (\ref{1Ds(4.3C)})) and $\tilde{\delta}(\omega)$, which are associated with flat-ended indentation tests, can elicit perceptions of the mechanical quality of articular cartilage. In the dynamic non-destructive testing with a spherical indenter, in view of the fact that the thickness of a biological tissue sample is supposed to be unknown, the pronounced effect of the sample thickness on the modified loss angle $\tilde{\delta}_{3/2}^0(\omega)$ observed for different levels of indentation can be used as an indicator of the importance of the thickness effect for the modified storage modulus $\tilde{E}_{3/2}^0(\omega,\varpi_0)$. 

\section*{Acknowledgements}
One of the authors (I.A.) gratefully acknowledges the support from the European Union Seventh Framework Programme under contract number PIIF-GA-2009-253055.

\end{document}